\documentclass[twocolumn]{autart}    

\usepackage{graphicx}          
                               
\usepackage{outlines} 
\usepackage{amsmath} 
\usepackage{amssymb}  
\usepackage{nicefrac}
\usepackage{mathtools}

\usepackage[T1]{fontenc}

\usepackage{hyperref}

\usepackage{color}

\newif\ifarXiv
\arXivtrue

\begin{document}

\begin{frontmatter}

\title{Nonlinear Functional Estimation: \\ Functional Detectability and Full Information Estimation\thanksref{footnoteinfo}} 

\thanks[footnoteinfo]{The work of Simon Muntwiler was supported by the Bosch Research Foundation im Stifterverband. 
	This work has been supported by the Swiss National Science Foundation under NCCR Automation (grant agreement 51NF40\_180545).
	This paper was not presented at any IFAC 
meeting. Corresponding author: Simon Muntwiler.}

\author{Simon Muntwiler}\ead{simonmu@ethz.ch},         
\author{Johannes K\"ohler}\ead{jkoehle@ethz.ch},       
\author{Melanie N. Zeilinger}\ead{mzeilinger@ethz.ch}  

\address{Institute for Dynamic Systems and Control, ETH Z\"urich, Z\"urich, Switzerland}  


\begin{abstract}                          
We consider the design of functional estimators, i.e., approaches to compute an estimate of a nonlinear function of the state of a general nonlinear dynamical system subject to process noise based on noisy output measurements.
To this end, we introduce a novel functional detectability notion in the form of incremental input/output-to-output stability ($\delta$-IOOS).
We show that $\delta$-IOOS is a necessary condition for the existence of a functional estimator satisfying an input-to-output type stability property.
Additionally, we prove that a system is functional detectable if and only if it admits a corresponding $\delta$-IOOS Lyapunov function.
Furthermore, $\delta$-IOOS is shown to be a sufficient condition for the design of a stable functional estimator by introducing the design of a full information estimation (FIE) approach for functional estimation.
Together, we present a unified framework to study functional estimation with a detectability condition, which is necessary and sufficient for the existence of a stable functional estimator, and a corresponding functional estimator design.
The practical need for and applicability of the proposed functional estimator design is illustrated with a numerical example of a power system. 
\end{abstract}

\end{frontmatter}

\section{Introduction}

In many practical applications, information about a dynamical system can only be inferred through noisy output measurements.
State estimation is a common framework to estimate the information (state) of a dynamical system based on such noisy output measurements, see, e.g.,~\cite{Luenberger1971}.
In recent years, necessary and sufficient conditions for the design of a \emph{robustly} stable state estimator have been derived in the form of a suitable detectability condition~\cite{Allan2021,Allan2021a}.
However, in many applications, the full state of a nonlinear system is \emph{not} detectable, and hence \emph{no} stable state estimator exists.
Instead, often only a part of the system state may be detectable.
Such problems can be approached using functional estimation, where instead of an estimate of the full state only an estimate of a (typically lower dimensional) function of the system state is obtained, employing a so-called \emph{functional estimator}.
Applications of functional estimation include, e.g., fault detection~\cite{Emami2015}, average estimators~\cite{Niazi2020}, partial state observers~\cite{Trinh2006}, and state-norm-estimators~\cite{Cai2008,Muller2012}, see Remark~\ref{rem:examples_functional_estimation} below for more details.
In this paper, we study nonlinear functional estimation as a general framework to address estimation problems even if the full system state is not detectable.
\subsection{Related Work}\label{sec:related_work}
Functional estimators were introduced in~\cite[Sec.~IV]{Luenberger1971} for linear systems to estimate a linear functional of the state with a minimal order observer to reduce the computational requirements compared to a full state observer.
This concept was extended to estimate a vector-valued function of the systems state (see, e.g.,~\cite{Fairman1980,Tsui1985}); initially termed \textit{multi-functional} estimators, but later also referred to as \textit{functional estimators}.
In~\cite{Darouach2000,Darouach2020}, necessary and sufficient conditions for the design of a linear functional estimator to compute an estimate of a linear function of the state of a linear system with minimal order, i.e., the same order as the dimension of the function to be estimated, were introduced.
To consider the case where a minimal order estimator does not exist, the concept of functional observability and detectabilty for linear systems was introduced in~\cite{Fernando2010}, compare also~\cite{Darouach2022} and references therein.
Thereby, functional detectability is a direct generalization of the classical detectability of the state of a linear system~\cite{Kalman1960}, i.e., all modes are either asymptotically stable, observable, or orthogonal to the linear function to be estimated.
Functional observability notions for nonlinear systems were introduced in~\cite{Alamir2021,Montanari2022}.
In~\cite{Fernando2010}, a design of a reduced-order functional observer for linear systems was introduced assuming functional detectability, while in~\cite{Darouach2022}, general necessary and sufficient conditions for the existence of a reduced-order linear functional estimator for linear systems were presented.
Applications of functional estimation to special classes of nonlinear systems have been investigated in~\cite{Trinh2006,Venkateswaran2022,Wang2021}.
Overall, the design of functional estimators for general nonlinear and time-varying systems subject to process and measurement noise is still an open question.
In particular, necessary and sufficient conditions for the existence of such a general nonlinear functional estimator have not been provided to date.
\subsection{Contribution}
In this paper, we provide a general analysis and design framework for nonlinear functional estimation.
In particular, we study general nonlinear time-varying systems subject to process and measurement noise with the objective to estimate a nonlinear function of the system state from output measurements (Section~\ref{sec:problem_setup}).
On a technical level, the derived theory extends recent developments from~\cite{Allan2021,Allan2021a,knuefer2021MHE,Schiller2022}, where necessary and sufficient conditions for the existence of a robustly stable \emph{state} estimator are developed.
These works characterized detectability of the system state using an incremental input/output-to-state-stability ($\delta$-IOSS) Lyapunov function and designed a corresponding state estimator using full information estimation (FIE).
The analysis of functional estimation poses additional challenges since the resulting Lyapunov function is in general only positive semi-definite, and thus lacks (exponential) contraction.

In Section~\ref{sec:functional_estimation}, we introduce incremental input/output-to-output stability ($\delta$-IOOS) as a notion of functional detectability (Definition~\ref{def:dIOOS}) and show that it is necessary and sufficient for the existence of an incrementally input-to-output stable ($\delta$-IOS) functional estimator (Proposition~\ref{prop:dIOOS_necessary}).
Additionally, we show that a system is $\delta$-IOOS if and only if a corresponding $\delta$-IOOS Lyapunov function exists (Proposition~\ref{prop:dIOOS_Lyap}).
Furthermore, we present an FIE design for functional estimation (Section~\ref{sec:FIE}).
Thereby, the functional estimate at each time step is obtained as the solution to a nonlinear optimization problem fitting a sequence of estimates to all currently available output measurements.
Assuming the system is functional detectable, the considered FIE formulation (Section~\ref{sec:FIE_fromulation}) is proven to be $\delta$-IOS (Theorem~\ref{thm:FIFE_stability}).
In combination, Corollary~\ref{cor:sufficient_condition} yields the key result of this paper: functional detectability is necessary and sufficient for the existence of a stable functional estimator.
Given a stricter \textit{exponential} detectability condition, we provide a simpler and more intuitive FIE design using a quadratic objective function (Section~\ref{sec:FIE_quadratic_objective}).
In Section~\ref{sec:discussion}, we compare our result with existing results in the domain of functional estimation, state-norm estimation, and state estimation.
In particular, we show that the existing necessary and sufficient conditions for linear functional estimation presented in~\cite{Darouach2022} imply the existence of a simple quadratic $\delta$-IOOS Lyapunov function (Section~\ref{sec:discussion_functional}).
Finally, we demonstrate the practical applicability of the presented functional estimator design for an academic example, where we estimate the total power load in a power system which is not detectable (Section~\ref{sec:numerical_example}).
\subsection{Notation}
Let the non-negative real numbers be denoted by $\mathbb{R}_{\geq 0}$, the set of integers by $\mathbb{I}$, the set of all integers greater than or equal to $a$ for some $a \in \mathbb{R}$ by $\mathbb{I}_{\geq a}$, and the set of integers in the interval $[a,b]$ for some $a,b\in\mathbb{R}$ with $a\le b$ by $\mathbb{I}_{[a,b]}$.
Let $\|x\|$ denote the Euclidean norm of the vector $x \in \mathbb{R}^n$.
The quadratic norm with respect to a positive definite matrix $Q=Q^\top$ is denoted by $\|x\|_Q^2=x^\top Q x$, and the minimal and maximal eigenvalues of $Q$ are denoted by $\lambda_{\min}(Q)$ and $\lambda_{\max}(Q)$, respectively.
The identity matrix is denoted by $I_n\in\mathbb{R}^{n\times n}$. 
A function $\alpha:\mathbb{R}_{\geq 0}\rightarrow\mathbb{R}_{\geq 0}$ is of class $\mathcal{K}$ if it is continuous, strictly increasing, and satisfies $\alpha(0)=0$. 
If $\alpha$ is additionally unbounded, it is of class $\mathcal{K}_{\infty}$.
We denote the class of functions $\theta:\mathbb{I}_{\geq 0}\rightarrow \mathbb{R}_{\geq 0}$ that are continuous, non-increasing, and satisfy $\lim_{t\rightarrow\infty}\theta(t)=0$ by $\mathcal{L}$.
By $\mathcal{KL}$, we denote the functions $\beta:\mathbb{R}_{\geq 0}\times\mathbb{I}_{\geq 0}\rightarrow\mathbb{R}_{\geq 0}$ with $\beta(\cdot,t)\in\mathcal{K}$ and $\beta(r,\cdot)\in\mathcal{L}$ for any fixed $t\in\mathbb{I}_{\geq 0}$, $r\in\mathbb{R}_{\geq 0}$.
The $i$-th element of a vector $x\in\mathbb{R}^n$ is denoted as $[x]_i$ for $i \in \mathbb{I}_{[1,n]}$.

\section{Problem Setup}\label{sec:problem_setup}

We consider a nonlinear time-varying discrete-time system
\begin{subequations}
	\label{eq:sys}
	\begin{align}
	\label{eq:sys_1}
	x_{t+1}&=f(x_t,w_t,t),\\
	\label{eq:sys_2}
	y_t&=h(x_t,w_t,t),  \\
	\label{eq:sys_3}
	z_t& = \phi(x_t),
	\end{align}
\end{subequations}
where $t\in\mathbb{I}_{\geq 0}$ is the discrete time step, $x_t\in\mathbb{R}^{n_{\mathrm{x}}}$ is the system state, $w_t\in\mathbb{R}^{n_{\mathrm{w}}}$ the process and measurement noise, $y_t\in\mathbb{Y}=\mathbb{R}^{n_{\mathrm{y}}}$ a measured output, and $z_t\in\mathbb{Z} = \mathbb{R}^{n_{\mathrm{z}}}$ a virtual output that we cannot measure, but would like to estimate.
The choice of time-varying dynamics~\eqref{eq:sys_1} and measurements~\eqref{eq:sys_2} also allows us to cover the important case of non-autonomous systems driven by a control input.
Note that $w_t$ appears in the dynamics~\eqref{eq:sys_1} and measurement model~\eqref{eq:sys_2} and hence the setting includes separated process disturbance and measurement noise as a special case.

Noisy measurements $y_t$ can be obtained from sensors, while the virtual output $z_t$ cannot be measured.
Given some estimate $\bar{x}_{t_0}$ of the initial state $x_{t_0}$ of system~\eqref{eq:sys} at the initial time step $t_0\in\mathbb{I}_{\ge 0}$, the objective is to obtain an estimate~$\hat{z}_t$ of the virtual output $z_t$ at each time step $t\in\mathbb{I}_{\ge t_0}$.
Since the virtual output~$z_t$ is a function of the system state $x_t$, this objective can be achieved by the use of a functional estimator (see Section~\ref{sec:functional_estimator}).
Functional estimation has a multitude of applications depending on the choice of the function~$\phi(x_t)$ in~\eqref{eq:sys_3}.
Some examples are highlighted in the following remark.

\begin{rem}[Applications of functional estimation]\label{rem:examples_functional_estimation}
	\phantom{A}\vspace{-7mm}
	\begin{itemize}
		\item \textbf{Partial state estimation:} In case the full system state is not detectable, a functional estimator can be employed to estimate a part of the state by choosing $\phi(x_t)= Lx_t$, where $L$ is  a linear mapping to a subspace of the state.
		In~\cite{Trinh2006}, such a partial state estimation approach for a class of nonlinear systems is introduced. In~\cite{Niazi2020}, partial state estimation is applied to estimate the average state within clusters in a networked system.
		In Section~\ref{sec:numerical_example}, we apply our presented FIE approach for partial state estimation in a power system.
		\item \textbf{Combined state and parameter estimation:} 
		Often, the uncertain state $x_t$ can be decomposed as $x_t=[s_t^\top\ \theta^\top]^\top$, where $\theta$ is a constant unknown parameter. 
		Then, the design of a robustly stable state estimator requires both, detectability of the partial state $s_t$ and a persistency of excitation condition~\cite{Sui2011}.
		In the absence of persistency of excitation, it might still be possible to employ a functional estimator with $\phi(x_t)=s_t$ to get a reliable estimate of the partial state $s_t$,~cf.,~e.g.,~\cite{Muntwiler2023b}.
		\item \textbf{Output-feedback:}
		In case an appropriate state feedback $u_t=\kappa(x_t)$ is available, but the system state $x_t$ is not detectable from the available output measurements, it might still be possible to apply a functional estimator to directly estimate the control input to apply to the system by choosing $\phi(x_t)=\kappa(x_t)$.
		This problem is studied in~\cite{Luenberger1971} for a linear feedback $\phi(x_t) = Kx_t$.
		\item \textbf{Fault detection:} Frequently, faults in a system are detected by analyzing observer residuals, which can be posed as a functional estimation problem with $\phi(x_t)=y_t$~\cite{Emami2015}.
		In particular, for linear time-invariant systems functional detectability is trivially satisfied in this case~\cite[Lem.~3.1]{Emami2015}.
		The general functional estimation framework enables the application of fault detection also based on virtual output signals $\phi(x_t)$ that cannot be measured directly, as, e.g., done in~\cite{teixeira2015strategic}.
		\item \textbf{State-norm estimation:} Sometimes, it is sufficient to obtain an estimate of the norm of the state to monitor or control a system, which can be posed as estimating $\phi(x_t) = \|x_t\|$, see, e.g.,~\cite{Cai2008,Muller2012}.
		In Section~\ref{sec:state_norm_estimation}, we show that functional detectability ($\delta$-IOOS) implies the necessary and sufficient condition for the existence of a classical state-norm estimator.
	\end{itemize}
\end{rem}

We consider the general case, where we might also have system information in the form of constraints
\begin{align}\label{eq:constraints}
	x_t\in\mathbb{X},\ w_t\in\mathbb{W},\ \forall t\in\mathbb{I}_{\ge 0},
\end{align}
with $\mathbb{X}\subseteq\mathbb{R}^{n_{\mathrm{x}}}$ and $\mathbb{W}\subseteq\mathbb{R}^{n_{\mathrm{w}}}$.
Note that this is not restrictive, since we can always choose $\mathbb{X}=\mathbb{R}^{n_{\mathrm{x}}}$ and $\mathbb{W}=\mathbb{R}^{n_{\mathrm{w}}}$ in case no additional system knowledge is available.
Constraints of the form~\eqref{eq:constraints} can be used to include additional physical prior knowledge, such as, e.g., non-negativity of certain states or bounded noise, into an estimator.
In the following, we assume that the true sequences of states and noises of system~\eqref{eq:sys} always satisfy the constraints~\eqref{eq:constraints}.

We call any sequences $\mathbf{x}=\{x_t\}_{t=t_0}^{t_{\mathrm{end}}}$, $\mathbf{w}=\{w_t\}_{t=t_0}^{t_{\mathrm{end}}-1}$, 
$\mathbf{y}=\{y_t\}_{t=t_0}^{t_{\mathrm{end}}-1}$, and  $\mathbf{z}=\{z_t\}_{t=t_0}^{t_{\mathrm{end}}}$ satisfying~\eqref{eq:sys}-\eqref{eq:constraints} a solution of system~\eqref{eq:sys}-\eqref{eq:constraints} starting at some time step $t_0\in\mathbb{I}_{\ge 0}$ and ending at time step $t_{\mathrm{end}}\in\mathbb{I}_{\ge t_0}$.
The set of such solutions is denoted by
\begin{align}\label{eq:solution}
\{\mathbf{x},\mathbf{w},\mathbf{y},\mathbf{z}\} \in \Sigma_{t_0}^{t_{\mathrm{end}}} \subseteq \mathbb{X}^{T+1}\times\mathbb{W}^{T}\times\mathbb{Y}^{T}\times\mathbb{Z}^{T+1},
\end{align}
with length\footnote{Note that for $t_{\mathrm{end}} = t_0$, i.e., $T=0$, the solution consists of one single element $x_{t_0} \in \mathbb{X}$ and the corresponding virtual output~\eqref{eq:sys_3}.} $T = t_{\mathrm{end}} - t_0$.

\section{Functional Estimation and Detectability}\label{sec:functional_estimation}
In this section, we extend results from~\cite{Allan2021,allan2019lyapunov} to show that functional detectability is necessary for the existence of a stable functional estimator (Proposition~\ref{prop:dIOOS_necessary}) and that it holds if and only if a corresponding Lyapunov function exists (Proposition~\ref{prop:dIOOS_Lyap}).
In contrast to~\cite{Allan2021,allan2019lyapunov}, we consider positive semi-definite Lyapunov functions to generalize state estimation to functional estimation.

\subsection{Functional Estimator}\label{sec:functional_estimator}
A functional estimator computes an estimate of~\eqref{eq:sys_3} at each time step $t$ using available quantities, i.e., an estimate of the initial state $\bar{x}_{t_0}$, and estimates of the process and measurement noise $\{\bar{w}_j\}_{j=t_0}^{t-1}$ and output measurements $\{\bar{y}_j\}_{j=t_0}^{t-1}$ obtained up to the current time step.
\begin{defn}[Functional estimator]\label{def:functional_estimator}
	A functional estimator is a sequence of functions $\Psi_t:\mathbb{X}\times\mathbb{W}^{t-t_0}\times\mathbb{Y}^{t-t_0}\times\mathbb{I}_{\ge 0}\rightarrow\mathbb{R}^{n_{\mathrm{z}}}$ to compute estimates of the virtual output $z_t$~\eqref{eq:sys_3} at each time step $t\in\mathbb{I}_{\ge t_0}$ as
	\begin{align}\label{eq:functional_estimator}
	\hat{z}_t = \Psi_t\left(\bar{x}_{t_0},\{\bar{w}_j\}_{j=t_0}^{t-1},\{\bar{y}_j\}_{j=t_0}^{t-1},t_0\right),
	\end{align}
	where $\bar{w}_j$ and $\bar{y}_j$ are estimates of the noise $w_j$ and measurements $y_j$ at time step $j$, respectively. 
\end{defn}

In a typical application, the noise estimates $\bar{w}_j$ are chosen to be zero and the output measurement $\bar{y}_j$ equal to the noisy output measurement $y_j$ according to~\eqref{eq:sys_2}.
The more general definition provided here will be crucial to provide a consistent stability definition, similar to~\cite{Allan2021}.
In line with the nomenclature in related work (cf. Section~\ref{sec:related_work}), we call this a \textit{functional estimator}, even though $z_t$ is in general vector-valued.

\subsection{Input-to-Output Stability}\label{sec:input-to-output_stability}

For the functional estimator according to Definition~\ref{def:functional_estimator}, we consider the following notion of stability.

\begin{defn}[Input-to-output stability] \label{def:input_output_stability}
	The functional estimator~\eqref{eq:functional_estimator} is incrementally input-to-output stable ($\delta$-IOS) with the noise $w_j$ and output measurements $y_j$ as well as their estimates $\bar{w}_j$ and $\bar{y}_j$ as inputs and the virtual outputs $z_t, \hat{z}_t$ as outputs if there exist $\beta_1,\ \beta_2,\ \beta_3\in\mathcal{KL}$ such that
	\begin{align}
	\|z_t - \hat{z}_t\| \le& \max\left\{\beta_1(\|x_{t_0} - \bar{x}_{t_0}\|,t-t_0), \vphantom{\max_{j\in\mathbb{I}_{[0,t-1]}}} \right. \nonumber \\
	&\max_{j\in\mathbb{I}_{[t_0,t-1]}}\beta_2(\|w_j - \bar{w}_j\|,t-j-1), \label{eq:stability_max_form} \\
	&\left. \max_{j\in\mathbb{I}_{[t_0,t-1]}}\beta_3(\|y_j - \bar{y}_j\|,t-j-1)\right\}, \nonumber
	\end{align}
	for all $t_0\in\mathbb{I}_{\ge 0}$, $t\in\mathbb{I}_{\ge t_0}$, $\{\mathbf{x},\mathbf{w},\mathbf{y},\mathbf{z}\} \in \Sigma_{t_0}^{t}$, $\{\bar{w}_j\}_{j=t_0}^{t-1}\in\mathbb{W}^{t-t_0}$, $\{\bar{y}_j\}_{j=t_0}^{t-1}\in\mathbb{Y}^{t-t_0}$, and $\hat{z}_t$ according to~\eqref{eq:functional_estimator}.
	The functional estimator~\eqref{eq:functional_estimator} is exponentially incrementally input-to-output stable if additionally $\beta_1(r,t)=C_1\lambda_1^tr$, $\beta_2(r,t)=C_2\lambda_2^tr$, and $\beta_3(r,t)=C_3\lambda_3^tr$ with $\lambda_1, \lambda_2, \lambda_3 \in [0,1)$ and $C_1, C_2, C_3 > 0$.
\end{defn}

Definition~\ref{def:input_output_stability} describes incremental input-to-output stability of a functional estimator according to Definition~\ref{def:functional_estimator}.
Thereby, the inputs are the estimated and actual noise and the estimated and actual measured outputs, while the outputs are the virtual output and the corresponding functional estimate resulting from~\eqref{eq:functional_estimator}.

\subsection{Functional Detectability}\label{sec:functional_detecatbility}

In the following, we present an appropriate definition of functional detectability.
Additionally, we show that the presented notion of functional detectability is a necessary condition for the existence of a stable functional estimator according to Definition~\ref{def:input_output_stability}.
\begin{defn}[Functional detectability] \label{def:dIOOS}
	System~\eqref{eq:sys} is incrementally input/output-to-output-stable ($\delta$-IOOS) if there exist $\beta_{\mathrm{x}},\ \beta_{\mathrm{w}},\ \beta_{\mathrm{y}} \in \mathcal{KL}$ such that
	\begin{align}
	\|z_t - \tilde{z}_t \| \le& \max\left\{\beta_{\mathrm{x}}(\|x_{t_0} - \tilde{x}_{t_0}\|,t-t_0), \vphantom{\max_{j\in\mathbb{I}_{[0,t-1]}}} \right. \nonumber \\ 
	&\max_{j\in\mathbb{I}_{[t_0,t-1]}}\beta_{\mathrm{w}}(\|w_j-\tilde{w}_j\|,t-j-1), \label{eq:dIOOS} \\
	&\left.\max_{j\in\mathbb{I}_{[t_0,t-1]}}\beta_{\mathrm{y}}(\|y_j-\tilde{y}_j\|,t-j-1)\right\}, \nonumber
	\end{align}
	for all\hspace{1mm}$t_0\in\mathbb{I}_{\ge 0}$, $t\in\mathbb{I}_{\ge t_0}$, and any sequences $\{\mathbf{x},\mathbf{w},\mathbf{y},\mathbf{z}\} \in \Sigma_{t_0}^{t}$ and $\{\tilde{\mathbf{x}}, \tilde{\mathbf{w}}, \tilde{\mathbf{y}},\tilde{\mathbf{z}}\} \in \Sigma_{t_0}^{t}$.
	The system~\eqref{eq:sys} is exponentially $\delta$-IOOS if additionally $\beta_{\mathrm{x}}(r,t)=C_{\mathrm{x}}\lambda_{\mathrm{x}}^tr$, $\beta_{\mathrm{w}}(r,t)=C_{\mathrm{w}}\lambda_{\mathrm{w}}^tr$, and $\beta_{\mathrm{y}}(r,t)=C_{\mathrm{y}}\lambda_{\mathrm{y}}^tr$, with $\lambda_{\mathrm{x}}, \lambda_{\mathrm{w}}, \lambda_{\mathrm{y}} \in [0,1)$ and $C_{\mathrm{x}}, C_{\mathrm{w}}, C_{\mathrm{y}} > 0$.
\end{defn}

Equation~\eqref{eq:dIOOS} implies that if two sequences starting at time step $t_0$ are subject to the same noise $w=\tilde{w}$ and result in the same output measurements $y=\tilde{y}$, then asymptotically their virtual outputs $z,\ \tilde{z}$ will converge to the same value.
This property is thus called functional detectability.

The following proposition establishes that $\delta$-IOOS is a necessary condition for the existence of a functional estimator satisfying Definition~\ref{def:input_output_stability}.
\begin{prop}\label{prop:dIOOS_necessary}
	A system~\eqref{eq:sys} admits an incrementally input-to-output stable functional estimator $\Psi_t$ as stated in Definition~\ref{def:input_output_stability} only if it is $\delta$-IOOS according to Definition~\ref{def:dIOOS}.
\end{prop}
The proof can be found in\ifarXiv~Appendix~\ref{app:proof_prop_dIOOS_nec}\fi.
It makes use of two arbitrary sequences satisfying~\eqref{eq:solution} and employs the definition of $\delta$-IOS (Definition~\ref{def:input_output_stability}) to show that~\eqref{eq:dIOOS} holds and, consequently, $\delta$-IOOS (Definition~\ref{def:dIOOS}) is necessary.

\subsection{$\delta$-IOOS Lyapunov Function}\label{sec:dIOOS}

In the following, we introduce a Lyapunov-type characterization of the $\delta$-IOOS property (Definition~\ref{def:dIOOS}).

\begin{defn}[$\delta$-IOOS Lyapunov function]
	\label{def:IOOS_Lyap}
	A function $W_\delta:\mathbb{R}^{n_{\mathrm{x}}}\times\mathbb{R}^{n_{\mathrm{x}}}\times\mathbb{I}_{\ge 0}\rightarrow\mathbb{R}_{\geq 0}$ is an (exponential-decrease) $\delta$-IOOS Lyapunov function if there exist $\alpha_1,\ \alpha_2\in\mathcal{K}_{\infty}$, $\sigma_{\mathrm{w}},\ \sigma_{\mathrm{y}}\in\mathcal{K}$, and $\eta\in[0,1)$ such that
	\begin{subequations}
		\label{eq:IOOS_Lyap}
		\begin{align}
		\label{eq:IOOS_Lyap_1}
		&\alpha_1(\|z-\tilde{z}\|)\leq W_\delta(x,\tilde{x},t)\leq \alpha_2(\|x-\tilde{x}\|),\\
		\label{eq:IOOS_Lyap_2}
		&W_\delta\left(f(x,w,t), f(\tilde{x},\tilde{w},t), t+1 \right) \leq \eta W_\delta(x,\tilde{x},t) \\
		& \qquad +\sigma_{\mathrm{w}}(\|w-\tilde{w}\|)+\sigma_{\mathrm{y}}(\|y-\tilde{y}\|), \nonumber
		\end{align}
	\end{subequations}
	for all $t\in\mathbb{I}_{\ge 0}$, $\{x,w,y,z\}\ \in\ \mathbb{X}\times\mathbb{W}\times\mathbb{Y}\times\mathbb{Z}$, and $\{\tilde{x},\tilde{w},\tilde{y},\tilde{z}\}\ \in\ \mathbb{X}\times\mathbb{W}\times\mathbb{Y}\times\mathbb{Z}$, where $y=h(x,w,t)$, $\tilde{y}=h(\tilde{x},\tilde{w},t)$, $z=\phi(x)$, and $\tilde{z}=\phi(\tilde{x})$.
	A function $W_\delta$ is an exponential $\delta$-IOOS Lyapunov function if additionally $\alpha_1,\ \alpha_2,\ \sigma_{\mathrm{w}},\ \sigma_{\mathrm{y}}$ are quadratic functions.
\end{defn}
Note that using an exponential-decrease Lyapunov function is without loss of generality, since the existence of a standard Lyapunov function implies the existence of an exponential-decrease Lyapunov function~\cite{Allan2021}.
The Lyapunov function $W_\delta$ is time-varying due to the time-varying setup~\eqref{eq:sys}.
In the following, we show that the existence of a $\delta$-IOOS Lyapunov function (Definition~\ref{def:IOOS_Lyap}) is a necessary and  sufficient condition for system~\eqref{eq:sys} to be $\delta$-IOOS (Definition~\ref{def:dIOOS}).

\begin{prop}\label{prop:dIOOS_Lyap}
	A system is $\delta$-IOOS according to Definition~\ref{def:dIOOS} if and only if it admits a $\delta$-IOOS Lyapunov function according to Definition~\ref{def:IOOS_Lyap}.
\end{prop}

The proof can be found in\ifarXiv~Appendix~\ref{app:proof_prop_dIOOS_Lyap}\fi.
To prove the sufficient direction,~\eqref{eq:IOOS_Lyap_2} is first applied $t-t_0$ times to two arbitrary sequences satisfying~\eqref{eq:solution}, and then the upper and lower bounds in~\eqref{eq:IOOS_Lyap_1} are applied to obtain a bound of the form~\eqref{eq:dIOOS}.
For the necessary direction, we define a candidate $W_\delta$ Lyapunov function and show that~\eqref{eq:IOOS_Lyap} holds for all $x, \tilde{x} \in \mathbb{X}$ and $t\in\mathbb{I}_{\ge 0}$ if a system is $\delta$-IOOS according to Definition~\ref{def:dIOOS}.
Note that it is also possible to show the existence of a \textit{continuous} $\delta$-IOOS Lyapunov function by assuming continuity of the system dynamics~\eqref{eq:sys} and modifying the candidate Lyapunov function in the proof, similar to~\cite[Thm.~3.5]{Allan2021}.
The following corollary shows that the necessary and sufficient condition also applies to the \textit{exponential} special case.

\begin{cor}\label{cor:exponendial_dIOOS}
	A system is exponentially $\delta$-IOOS according to Definition~\ref{def:dIOOS} if and only if it admits an exponential $\delta$-IOOS Lyapunov function according to Definition~\ref{def:IOOS_Lyap}.
\end{cor}

In summary, this section showed that functional detectability according to Definition~\ref{def:dIOOS} is a necessary condition for the existence of a stable functional estimator (Proposition~\ref{prop:dIOOS_necessary}).
Furthermore, a system admits a $\delta$-IOOS Lyapunov function if and only if it is functional detectable according to Definition~\ref{def:dIOOS}.
In the next section, we show that the presented detectability property is also a sufficient condition for the existence of a $\delta$-IOS functional estimator.
\section{Full Information Estimation}\label{sec:FIE}
In this section, we present an FIE formulation for systems admitting a $\delta$-IOOS Lyapunov function (Section~\ref{sec:FIE_fromulation}) and prove that it is a $\delta$-IOS functional estimator according to Definition~\ref{def:input_output_stability} (Section~\ref{sec:FIE_theoretical_analysis}).
This implies that functional detectability is not only necessary, but also sufficient for the existence of a $\delta$-IOS functional estimator (Corollary~\ref{cor:sufficient_condition}).
In Section~\ref{sec:FIE_quadratic_objective}, we additionally show how the corresponding FIE design simplifies in case of \textit{exponential} $\delta$-IOOS by using a quadratic objective.
\subsection{FIE Formulation}\label{sec:FIE_fromulation}
The presented FIE approach considers all available estimates of past output measurements $\left\{\bar{y}_j\right\}_{j=t_0}^{t-1}$ and noise $\left\{\bar{w}_j\right\}_{j=t_0}^{t-1}$, and the initial estimate $\bar{x}_{t_0}$ to obtain the current functional estimate $\hat{z}_t$ at time step $t$.
Thereby, the FIE optimizes over an initial estimate $\hat{x}_{t_0|t}$ and a sequence of noise estimates $\hat{w}_{\cdot | t} = \{\hat{w}_{j|t}\}_{j=t_0}^{t-1}$, which uniquely define a corresponding sequence of states $\hat{x}_{\cdot | t} = \{\hat{x}_{j|t}\}_{j=t_0}^{t-1}$ and outputs $\hat{y}_{\cdot | t} = \{\hat{y}_{j|t}\}_{j=t_0}^{t-1}$.
The FIE objective is given by
\begin{align}
V_{\mathrm{FIE}}(\hat{x}_{t_0|t},\hat{w}_{\cdot|t},&\hat{y}_{\cdot|t},t) \coloneqq \eta^{t}\alpha_2\left(2\|\hat{x}_{t_0|t}-\bar{x}_{t_0}\|\right) \nonumber \\
&+\sum_{j=1}^{t-t_0}\eta^{j-1}\sigma_{\mathrm{w}}(2\|\hat{w}_{t-j|t} - \bar{w}_{t-j}\|) \label{eq:FIE_objective} \\
&+\sum_{j=1}^{t-t_0}\eta^{j-1}\sigma_{\mathrm{y}}(2\|\hat{y}_{t-j|t}-\bar{y}_{t-j}\|), \nonumber
\end{align}
where $\eta,\ \alpha_2,\ \sigma_{\mathrm{w}},\ \sigma_{\mathrm{y}}$ are based on the $\delta$-IOOS Lyapunov function according to Definition~\ref{def:IOOS_Lyap}.
In Section~\ref{sec:FIE_quadratic_objective}, we discuss how this requirement simplifies in case of an exponential detectability condition.
The time-discounting in the FIE objective~\eqref{eq:FIE_objective} decreases the influence of old measurements on the current estimate and it allows us to relate the FIE objective to the functional detectability condition~\eqref{eq:dIOOS}.
A similar time-discounting in the objective for FIE has been used, e.g., in~\cite{knuefer2021MHE,Schiller2022} for state estimation, compare~\cite[Sec. III.D]{Schiller2022} for a discussion of the benefits of such a time-discounting.
The estimate at each time step~$t\in\mathbb{I}_{\geq t_0}$ is then obtained by solving the following nonlinear program
\begin{subequations}\label{eq:FIE_IOOS}
	\begin{alignat}{2}\label{eq:FIE_IOOS_cost}
	&\inf_{\hat{x}_{t_0|t},\hat{w}_{\cdot|t}} V_{\mathrm{FIE}}(\hat{x}_{t_0|t},\hat{w}_{\cdot|t},\hat{y}_{\cdot|t},t) \\ \label{eq:FIE_IOOS_1}
	&\hspace{0.75cm} \text{s.t. }\hat{x}_{j+1|t}=f(\hat{x}_{j|t},\hat{w}_{j|t},j),&& \quad j\in\mathbb{I}_{[t_0,t-1]},\\
	\label{eq:FIE_IOOS_2}
	&\hspace{1.3cm} \hat{y}_{j|t}=h(\hat{x}_{j|t},\hat{w}_{j|t},j),&&\quad j\in\mathbb{I}_{[t_0,t-1]},\\
	\label{eq:FIE_IOOS_3}
	&\hspace{1.3cm}\hat{w}_{j|t} \in\mathbb{W}, &&\quad j\in\mathbb{I}_{[t_0,t-1]}, \\
	\label{eq:FIE_IOOS_4}
	&\hspace{1.3cm}\hat{x}_{j|t}\in\mathbb{X}, &&\quad j\in\mathbb{I}_{[t_0,t]}.
	\end{alignat}
\end{subequations}
A (non-unique) minimizer\footnote{We assume that a minimizer to~\eqref{eq:FIE_IOOS} always exists.
Existence of such a minimizer can, e.g., be ensured if $\mathbb{X}$ and $\mathbb{W}$ are compact sets and $f\ \text{and}\ h$ are continuous functions (see, e.g.,~\cite[Prop. A.7]{Rawlings2020}).} to~\eqref{eq:FIE_IOOS} is denoted as $\hat{x}_{t_0|t}^*$ and $\hat{w}_{\cdot | t}^*$, with corresponding estimated state and output sequences denoted as $\hat{x}_{\cdot | t}^*$ and $\hat{y}_{\cdot | t}^*$, respectively.
Finally, the functional estimate at time step $t$ is denoted as
\begin{align}\label{eq:FIE_estimate}
\hat{z}_t = \phi(\hat{x}_{t|t}^*).
\end{align}
Using an optimization-based FIE formulation~\eqref{eq:FIE_IOOS} allows us to use the full nonlinear model in~\eqref{eq:FIE_IOOS_1}-\eqref{eq:FIE_IOOS_2} and to consider additional information in the form~\eqref{eq:constraints} by enforcing~\eqref{eq:FIE_IOOS_3}-\eqref{eq:FIE_IOOS_4}.
It follows that any minimizer to~\eqref{eq:FIE_IOOS} is a solution to~\eqref{eq:sys}-\eqref{eq:constraints} and the true sequence of system states (with corresponding sequences of noise and measurements) is a feasible solution to~\eqref{eq:FIE_IOOS}.
In the following, we investigate the theoretical properties of the presented FIE approach~\eqref{eq:FIE_IOOS} and the resulting functional estimate~\eqref{eq:FIE_estimate}.

\subsection{Theoretical Analysis}\label{sec:FIE_theoretical_analysis}
This section generalizes results from~\cite{knuefer2021MHE,Schiller2022} to the case of functional estimation with semi-definite Lyapunov functions.
The following theorem summarizes the stability properties of the FIE approach~\eqref{eq:FIE_IOOS}.

\begin{thm}\label{thm:FIFE_stability}
	Let system~\eqref{eq:sys} admit a $\delta$-IOOS Lyapunov function according to Definition~\ref{def:IOOS_Lyap}. Then, the FIE~\eqref{eq:FIE_IOOS} is $\delta$-IOS according to Definition~\ref{def:input_output_stability}.
\end{thm}

The proof can be found in \ifarXiv Appendix~\ref{app:proof_thm_FIE_stability}\fi.
It relies on the fact that the optimal objective~\eqref{eq:FIE_IOOS_cost} is upper-bounded by the objective~\eqref{eq:FIE_objective} with the true, and hence feasible, sequences.
Consequently, a relationship between the FIE objective~\eqref{eq:FIE_objective} and the $\delta$-IOOS Lyapunov function according to Definition~\ref{def:IOOS_Lyap} can be established, which allows us to obtain a bound of the form~\eqref{eq:stability_max_form}.
The following corollary demonstrates that $\delta$-IOOS is a necessary and sufficient condition for the existence of a $\delta$-IOS functional estimator.
\begin{cor}\label{cor:sufficient_condition}
		A system~\eqref{eq:sys} admits a $\delta$-IOS functional estimator $\Psi_t$ as stated in Definition~\ref{def:input_output_stability} if and only if it is $\delta$-IOOS according to Definition~\ref{def:dIOOS}.
\end{cor}
\begin{pf}
	\textbf{Sufficiency:} As a result of Proposition~\ref{prop:dIOOS_Lyap}, the system admits a $\delta$-IOOS Lyapunov function according to Definition~\ref{def:IOOS_Lyap}, if it is $\delta$-IOOS according to Definition~\ref{def:dIOOS}.
	Theorem~\ref{thm:FIFE_stability} showed that the FIE~\eqref{eq:FIE_IOOS} is a $\delta$-IOS functional estimator according to Definition~\ref{def:input_output_stability}, if $\eta,\ \alpha_2,\ \sigma_{\mathrm{w}},\ \sigma_{\mathrm{y}}$ are based on a $\delta$-IOOS Lyapunov function.
	In combination, this shows that $\delta$-IOOS is sufficient for the existence of a $\delta$-IOS functional estimator. \\
	\textbf{Necessity:} It follows directly from Proposition~\ref{prop:dIOOS_necessary} that $\delta$-IOOS is necessary for the existence of a $\delta$-IOS functional estimator, which concludes the proof. \vspace*{-0.8cm}\flushright\qed
\end{pf}

We introduced an FIE design based on $\delta$-IOOS, which results in a $\delta$-IOS functional estimator (Theorem~\ref{thm:FIFE_stability}) and shows that $\delta$-IOOS is a necessary and sufficient condition for the existence of a $\delta$-IOS functional estimator (Corollary~\ref{cor:sufficient_condition}).
However, in general it is difficult to explicitly derive a corresponding $\delta$-IOOS Lyapunov function $W_\delta$ to obtain $\eta$, $\alpha_2$, $\sigma_{\mathrm{w}}$, and $\sigma_{\mathrm{y}}$ for the design of the FIE objective~\eqref{eq:FIE_objective}.
In the following, we thus present a simplified design for the special case of \textit{exponentially} functional detectable systems according to Definition~\ref{def:dIOOS}.

\subsection{FIE with Quadratic Objective}\label{sec:FIE_quadratic_objective}
For the case of \emph{exponentially} functional detectable systems (Definition~\ref{def:dIOOS}), we use an FIE formulation with general quadratic objective and show that the resulting FIE approach is exponentially $\delta$-IOS. 
We consider the following FIE objective
\begin{align}
&V_{\mathrm{FIE}}(\hat{x}_{t_0|t},\hat{w}_{\cdot|t},\hat{y}_{\cdot|t},t) \coloneqq 2\eta^{t}\|\hat{x}_{t_0|t}-\bar{x}_{t_0}\|_{P}^2 \label{eq:FIE_objective_quad} \\
&\quad +2\sum_{j=1}^{t-t_0}\eta^{j-1}\left(\|\hat{w}_{t-j|t}-\bar{w}_{t-j}\|_{Q}^2+\|\hat{y}_{t-j|t}-\bar{y}_{t-j}\|_{R}^2 \right), \nonumber
\end{align}
with $\eta \in [0,1)$ and arbitrary matrices $P,\ Q,\ R \succ 0$.
The FIE approach is then applied in the same manner as the general formulation introduced in Section~\ref{sec:FIE_fromulation}.
Specifically, the FIE problem~\eqref{eq:FIE_IOOS} with objective~\eqref{eq:FIE_objective_quad} is solved at each time step $t\in\mathbb{I}_{\ge t_0}$ using all available noise estimates $\bar{w}$ and measurement estimates $\bar{y}$, and the resulting functional estimate is obtained using~\eqref{eq:FIE_estimate}.
The following corollary establishes that we can use any positive definite weighting matrices in~\eqref{eq:FIE_objective_quad}, provided that the decay factor $\eta<1$ is chosen sufficiently close to $1$. 
\begin{cor}\label{cor:FIE_quad_cost}
	Let system~\eqref{eq:sys} be exponentially $\delta$-IOOS according to Definition~\ref{def:dIOOS}.
	Suppose further that $1 > \eta \ge \max\{\lambda_{\mathrm{x}},\ \lambda_{\mathrm{w}},\ \lambda_{\mathrm{y}}\}$ and $P,\ Q,\ R\succ 0$ are arbitrary positive definite matrices.
	Then, the FIE~\eqref{eq:FIE_IOOS} with objective~\eqref{eq:FIE_objective_quad} is exponentially $\delta$-IOS according to Definition~\ref{def:input_output_stability}. 
\end{cor}
The proof can be found in\ifarXiv~Appendix~\ref{app:proof_cor_FIE_quad_cost}\fi.
This result provides great flexibility requiring only minimal system knowledge ($\eta$ sufficiently close to $1$) when designing the FIE objective~\eqref{eq:FIE_objective_quad} for systems~\eqref{eq:sys} that are exponentially $\delta$-IOOS according to Definition~\ref{def:dIOOS}.
Provided the decay factor $\eta$ is close enough to $1$ (depending on functional detectability of the system), any choice of positive definite weighting matrices results in a $\delta$-IOS FIE approach.
Hence, the weighting matrices can be freely chosen, e.g., based on the inverse covariance matrices.
In particular, in case the discount factor $\eta$ approaches one and neglecting the constraints~\eqref{eq:constraints}, the objective~\eqref{eq:FIE_objective_quad} can be interpreted as a maximum a posteriori estimate under Gaussian distributions, compare~\cite{Rao2000}.
\section{Discussion}\label{sec:discussion}
In this section, we provide a discussion of the relation of our results to existing work in the context of functional estimation (Section~\ref{sec:discussion_functional}), state-norm estimation (Section~\ref{sec:state_norm_estimation}), and standard state estimation (Section~\ref{sec:state_estimation}).
\subsection{Relation to Existing Functional Estimation Results}\label{sec:discussion_functional}
In the following, we contrast the presented conditions to existing results for functional estimation.
Most existing work on functional estimation considers linear systems, and we show that under the existing necessary and sufficient conditions for such linear functional estimators, we can construct a quadratic $\delta$-IOOS Lyapunov function.

We consider the special case of linear time-invariant systems
\begin{align}\label{eq:sys_lin}
x_{t+1} = Ax_t + Bw_t,\ y_t = Cx_t + Dw_t,\ z_t = Lx_t,
\end{align}
which is similarly considered in~\cite{Darouach2000,Fernando2010,Darouach2020,Darouach2022}\footnote{
In~\cite{Darouach2000,Fernando2010,Darouach2020,Darouach2022}, continuous-time systems are considered, but the results naturally extend to discrete-time systems subject to process and measurement noise.
We conjecture that the proof of the necessary and sufficient conditions presented in~\cite{Darouach2022} generalizes to the considered discrete-time setting.
}.
Therein, a reduced-order functional estimator of the form\footnote{\label{ft:current_measurement}In related work, the functional estimator~\eqref{eq:lin_estimator} often includes the current output measurement $y_t$, i.e., $\hat{z}_t = P_\xi	\xi_t + J_1	y_t$.
Consequently, Condition~\eqref{eq:lin_nec_cond_2} reads $P_\xi T + J_1C = L$.
We choose $J_1=0$ to ensure the functional estimator is strictly causal (cf. Definition~\ref{def:functional_estimator}).}
\begin{align}\label{eq:lin_estimator}
	\xi_{t+1} = N\xi_t + Jy_t,\ \hat{z}_t = P_\xi	\xi_t,\ \xi_0 \in\mathbb{R}^{n_\xi},
\end{align}
with reduced order $n_\xi \in\mathbb{I}_{[n_{\mathrm{z}},n_{\mathrm{x}}]}$ is designed.
In~\cite{Darouach2022}, the following necessary and sufficient condition for the existence of an asymptotically stable functional estimator of the form~\eqref{eq:lin_estimator} for system~\eqref{eq:sys_lin} (assuming $w_t=0$ for all $t\in\mathbb{I}_{\ge 0}$) was introduced: There exist matrices $N\in\mathbb{R}^{n_\xi \times n_\xi}$, $J\in\mathbb{R}^{n_{\mathrm{y}} \times n_\xi}$, $P_\xi\in\mathbb{R}^{n_\xi \times n_{\mathrm{z}}}$, and $T\in\mathbb{R}^{n_\xi \times n_{\mathrm{x}}}$ with $N$ Schur such that
\begin{subequations}\label{eq:lin_nec_cond}
	\begin{align}
	NT -TA + JC =&\ 0, \label{eq:lin_nec_cond_1} \\
	P_\xi T =&\ L. \label{eq:lin_nec_cond_2}
	\end{align}
\end{subequations}
In case Conditions~\eqref{eq:lin_nec_cond} is satisfied, one can show that system~\eqref{eq:sys_lin} admits a quadratic $\delta$-IOOS Lyapunov function according to Definition~\ref{def:IOOS_Lyap} of the form
\begin{align}\label{eq:dIOSS_lin}
W_\delta(x,\tilde{x}) = \| T(x-\tilde{x}) \|_P^2,
\end{align}
with $P\succ 0$, $P\in\mathbb{R}^{n_\xi \times n_\xi}$, such that
\begin{align} \label{eq:Lyap_eq}
	N^\top P N \preceq \rho P 
\end{align}
and $\rho < 1$, see\ifarXiv~Appendix~\ref{app:lin_functional}\fi~for a detailed proof.
Note that the estimator~\eqref{eq:lin_estimator} is of dimension $n_\xi$, which is in general larger than the dimension of $z_t$.
From Proposition~\ref{prop:dIOOS_Lyap}, it follows that the linear system~\eqref{eq:sys_lin} is $\delta$-IOOS according to Definition~\ref{def:dIOOS}.
Consequently, classical functional detectability~\cite{Fernando2010,Darouach2022} is recovered as a special case of $\delta$-IOOS.
A similar relation can be established for the results presented in~\cite{Trinh2006}, where necessary and sufficient conditions for the existence of a reduced order nonlinear functional estimator for partial state estimation were investigated for the special case where the functions $\phi$ and $h$ in~\eqref{eq:sys_2}-\eqref{eq:sys_3} are time-invariant and linear in $x_t$, i.e., $h(x_t,w_t,t)=C_{\mathrm{h}} x_t$ and $\phi(x_t)=C_\phi x_t$.

In contrast, this paper presents necessary and sufficient conditions for the existence of a more general class of functional estimators applicable to a more general class of nonlinear systems of the form~\eqref{eq:sys}.
In turn, the provided FIE design for functional estimation is more complex compared to reduced order observers as considered in~\cite{Darouach2000,Fernando2010,Darouach2020,Darouach2022}.
The design of reduced order observers for general nonlinear systems remains an interesting question for future research.
\subsection{Relation to State-Norm Estimation}\label{sec:state_norm_estimation}
A special case of functional estimation is state-norm estimation where $\phi(x_t) = \|x_t\|$, compare,~\cite{Cai2008,Muller2012}.
A system admits a state-norm estimator according to~\cite[Def.~3.11]{Cai2008} if it is input/output-to-state-stable (IOSS)~\cite[Def.~3.4]{Cai2008}.
In this case, a simple scalar state-norm estimator is given by
\begin{align}\label{eq:state_norm_estimator}
	\hat{z}_{t+1} = \epsilon \hat{z}_t + \rho_1(\|\bar{y}_t\|) + \rho_1(\|\bar{w}_t\|), 
\end{align}
where $\epsilon \in [0,1)$, and $\rho_1,\ \rho_2 \in \mathcal{K}$.
It follows directly\footnote{Assuming the origin is contained within $\mathbb{X}$ and $\mathbb{W}$, and w.l.o.g. that $0=f(0,0,t),\ 0=h(0,0,t)$ for all $t \in \mathbb{I}_{\ge 0}$.} from Definition~\ref{def:dIOOS}, that a system is IOSS if it is $\delta$-IOOS with $\phi(x_t) = \|x_t\|$, allowing for the design of a state-norm estimator of the form~\eqref{eq:state_norm_estimator} (as a consequence of~\cite[Thm.~3.13]{Cai2008}).
However, the definition of a state-norm estimator~\cite[Def.~3.11]{Cai2008} is significantly weaker compared to the stability properties considered in Definition~\ref{def:input_output_stability}.
In particular, the bounds in~\cite[Def.~3.11]{Cai2008} do \emph{not} ensure that the estimate of the state-norm $\hat{z}_t$ converges to the true state-norm $\|x_t\|$ in the absence of noise.
Therefore, IOSS does in general not imply $\delta$-IOOS with $\phi(x_t)=\|x_t\|$.

\subsection{Special Case --- State Estimation}\label{sec:state_estimation}
In the special case with $\phi(x_t)=x_t$, our theoretical analysis from Section~\ref{sec:functional_estimation} and~\ref{sec:FIE} recovers important results from classical state estimation, in particular~\cite[Prop. 2.6]{Allan2021} and~\cite[Thm. 3.2]{Allan2021}.
A general limitation of the FIE approach~\eqref{eq:FIE_IOOS} is the fact that the size of the optimization problem increases over time. 
In the context of state estimation, this limitation can be overcome by applying a moving horizon estimation approach (see, e.g.,~\cite{Rao2000,knuefer2021MHE,Schiller2022}), where only a finite horizon of past noise and output estimates are considered to obtain a state estimate $\hat{x}_t$ at each time step $t$.
However, depending on the choice of $\phi(x_t)$, it might still be possible to design an input-to-output stable functional estimator, even if it is \emph{not} possible to design a robustly stable state estimator, see Section~\ref{sec:numerical_example} for a numerical example.

\section{Numerical Example}\label{sec:numerical_example}
In this section, we apply the FIE approach for functional estimation as developed in Section~\ref{sec:FIE} to estimate the total power load in a nonlinear power system example as described in Section~\ref{sec:example_model}.
For this example, we show that the full system state is not detectable, while the total power load is functional detectable, allowing us to design a $\delta$-IOS functional estimator.
The numerical example was implemented in python using CasADi~\cite{Andersson2019} and the solver IPOPT~\cite{Waechter2005}.
The code\footnote{\href{https://doi.org/10.3929/ethz-b-000690515}{https://doi.org/10.3929/ethz-b-000690515}} is available online.
\subsection{Power System Model}\label{sec:example_model}
We consider a nonlinear power system model adapted from~\cite{Zhao2013,Li2016}.
In particular, we consider a system with a set of $4$ buses $\mathcal{N} = \{1,2,3,4\}$ and a set of $4$ transmission lines $(i,j)\in\mathcal{E} = \{(1,2),(2,3),(3,4),(4,1)\}$ connecting the buses $i$ and~$j$.
To monitor and control such a system, it is important to have an estimate of the power load within the system at each time step.
While the mechanical power generated at each bus is set by the system operator, and we thus assume to have measurements thereof available, the power load is affected by external factors and thus assumed not to be measured.

The nominal nonlinear system dynamics at each bus $i\in\mathcal{N}$ are described by
\begin{subequations}\label{eq:power_sys}
	\begin{align}
		[\dot{\theta}]_i =&\ [\omega]_i,\ [\dot{P}^L]_i = 0,\
			[\dot{P}^M]_i = 0, \\
		[\dot{\omega}]_i =&\ -\frac{1}{M_i} \left(D_i[\omega]_i-[P^M]_i+[P^L]_i+[\Delta P]_i\right), \label{eq:power_sys_freq} \\
		[\Delta P]_i =& \sum_{k:(i,k)\in\mathcal{E}}P_{ik}-\sum_{j:(j,i)\in\mathcal{E}}P_{ji}, \label{eq:delta_P_i} \\
		P_{ij} =& 3\frac{|V_i||V_j|}{x_{ij}}\sin\left(\theta_i - \theta_j\right),~\forall~(i,j)\in\mathcal{E}, \label{eq:branch_flow}
	\end{align}
\end{subequations}
where $\theta \in \mathbb{R}^{N}$ is the phase angle, $\omega \in \mathbb{R}^N$ the frequency deviation from a nominal frequency, $P^L \in \mathbb{R}^N$ the power load, $P^M \in \mathbb{R}^N$ the mechanical power, $\Delta P\in\mathbb{R}^{N}$ the power outflow, $P_{ij}$ the branch flow across each line, $n_{\mathrm{x}}=4\cdot N = 16$ the number of states, and $x = [\theta^\top\ \omega^\top\ {P^L}^\top\ {P^M}^\top ]^\top\in\mathbb{X}=\mathbb{R}^{n_{\mathrm{x}}}$ is the overall system state.
Additionally, $M_i$ is the generator inertia, $D_i$ the damping constant, and $|V_i|$ the voltage magnitude of each bus $i \in \mathcal{N}$, and $x_{ij}$ is the reactance of each line $(i,j)\in\mathcal{E}$.
The derivation of the branch flow model~\eqref{eq:branch_flow} can be found in~\cite[App. VII.A]{Zhao2013}\footnote{This power flow model assumes zero line resistance and constant voltage magnitudes.}.
The numerical values of the system parameters are chosen as in~\cite[Sec. VI.]{Li2016}.
The system dynamics~\eqref{eq:power_sys} are discretized using Euler forward with sampling time $\Delta t = 0.01$, and the resulting discrete-time model is subject to a uniformly distributed additive process noise $w_{\mathrm{x}}\in\mathbb{W}_{\mathrm{x}}=\{w_{\mathrm{x}}\in\mathbb{R}^{n_{\mathrm{x}}}|\|w_{\mathrm{x}}\|_\infty\leq 5\cdot 10^{-3}\}$.
At each time step, we obtain noisy measurements of the frequency $\omega$ and mechanical power $P^M$, i.e.,
\begin{align} \label{eq:example_meas}
	y = \begin{bmatrix}
	\omega \\
	P^M
	\end{bmatrix} + w_{\mathrm{y}} \in\mathbb{R}^{n_{\mathrm{y}}}, \quad n_{\mathrm{y}}=2\cdot N=8,
\end{align}
which are subject to uniformly distributed additive measurement noise $w_{\mathrm{y}}\in\mathbb{W}_{\mathrm{y}}=\{w_{\mathrm{y}}\in\mathbb{R}^{n_{\mathrm{y}}}|\ \| w_{\mathrm{y}}\|_\infty \le 5\cdot 10^{-2}\}$.
The overall process and measurement noise is denoted as $w=[w_{\mathrm{x}}^\top\ w_{\mathrm{y}}^\top]^\top\in\mathbb{R}^{n_{\mathrm{x}}+n_{\mathrm{y}}}$.
The objective is to estimate the total power load given by
\begin{align}\label{eq:example_total_load}
	z = \sum_{i\in\mathcal{N}}[P^L]_i.
\end{align}
\subsection{Functional Detectability}\label{sec:example_functional_detectability}
In this section, we establish that the state of system~\eqref{eq:power_sys} is not detectable from the available output measurements~\eqref{eq:example_meas}, while the system is functional detectable with the total load~\eqref{eq:example_total_load} as virtual output of interest.

To establish that the full system state is not detectable, we consider the noise-free case, i.e., $w=0$, and any steady-state of system~\eqref{eq:power_sys}.
Consequently, we have that $\omega = 0$ and $P^M = P^L + \Delta P$.
It follows that different combinations of power loads $P^L$ and phase angles $\theta$ lead to the same (measured) mechanical power $P^M$.
Since both $P^L$ and $\theta$ are not asymptotically stable, it follows that the full system state, and in particular the power load $[P^L]_i$ at each bus $i\in\mathcal{N}$, is \emph{not} detectable.

To establish functional detectability of the considered example, note that
\begin{align} \label{eq:linear_functional_estimator}
	z_t = C_{\mathrm{y}} \begin{bmatrix}
	y_{t-1} \\
	y_{t-2}
	\end{bmatrix} + C_{\mathrm{w}} \begin{bmatrix}
	w_{t-1} \\
	w_{t-2}
	\end{bmatrix},
\end{align}
with appropriate choice of matrices $C_{\mathrm{y}}\in\mathbb{R}^{1\times 2\cdot n_{\mathrm{y}}}$ and $C_{\mathrm{w}}\in\mathbb{R}^{1\times 2 \cdot (n_{\mathrm{x}}+n_{\mathrm{y}})}$, see\ifarXiv~Appendix~\ref{app:example_derivation}\fi~for a detailed derivation.
Therefore, the system is functional observable~\cite{Fernando2010}, and in the noise-free case ($w=0$), the functional value at time step $t$ can be exactly reconstructed given the past two measurements.
It follows that the system is \emph{exponentially} $\delta$-IOOS with any decay factor $\eta\in (0,1)$.
\subsection{Estimation Results}\label{sec:example_estimation}
\begin{figure}
	\centering
	\includegraphics[width=\columnwidth]{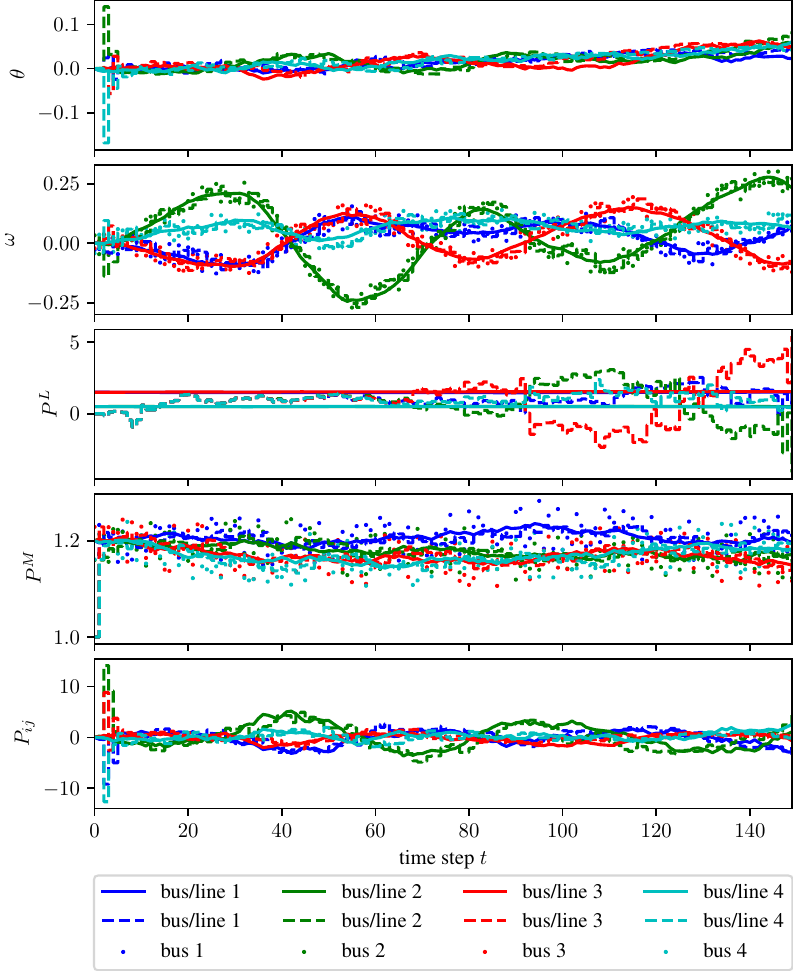}
	\caption{Ground truth states at each bus and branch flow across each line (solid lines) with corresponding measurements (dots) and FIE estimates (dashed lines).
	}
	\label{fig:power_system_states}
\end{figure}
\begin{figure}
	\centering
	\includegraphics[width=\columnwidth]{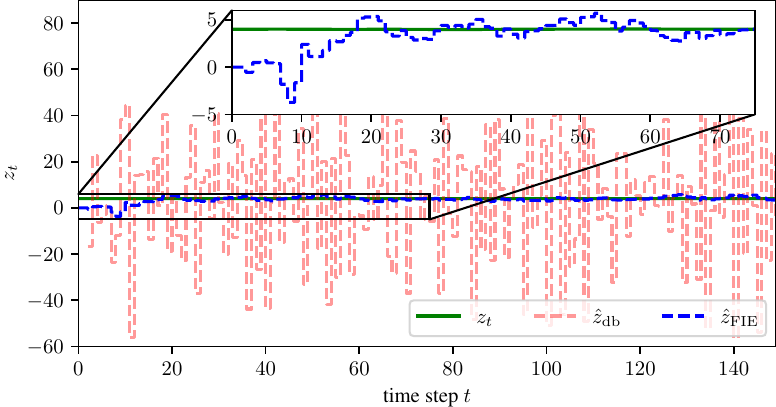}
	\caption{Functional value (solid green) and corresponding functional estimate resulting from our FIE approach (dashed blue) and from the simple linear estimator~\eqref{eq:linear_functional_estimator} (dashed red).}
	\label{fig:power_system_functional}
\end{figure}
\begin{figure}
	\centering
	\includegraphics[width=\columnwidth]{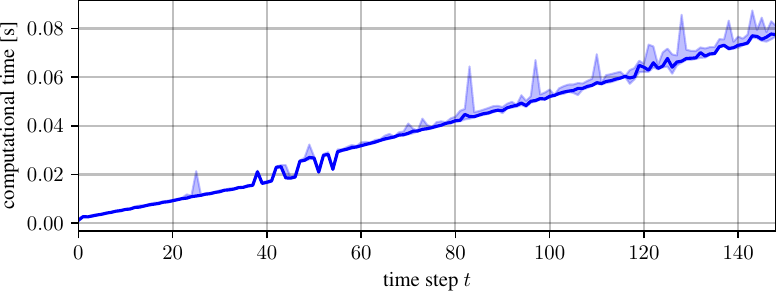}
	\caption{Median computational time required to solve the FIE problem depending on the time step for $10$ different runs. The shaded area shows the corresponding minimal and maximal computational time.}
	\label{fig:power_system_comp_time}
\end{figure}
To estimate the total power load~\eqref{eq:example_total_load} we design an FIE approach with quadratic objective~\eqref{eq:FIE_objective_quad} as outlined in Section~\ref{sec:FIE_quadratic_objective}.
We chose the decay factor  $\eta = 0.9$, and design the weighting matrices based on the inverse covariance matrices of process and measurements noise.
The ground truth system states (solid lines), and corresponding FIE state estimates (dashed lines) and measurements of $\omega$ and $P^M$ (dots) can be found in Figure~\ref{fig:power_system_states}.
In Figure~\ref{fig:power_system_functional}, we show the ground truth functional value and the functional estimate $\hat{z}_{\mathrm{FIE}}$ resulting from our FIE approach (blue).
For comparison, we also show the functional estimate $\hat{z}_{\mathrm{db}}$ resulting form a deadbeat estimator, i.e., using Equation~\eqref{eq:linear_functional_estimator} and neglecting the noise.
It can be seen that the FIE estimate quickly converges to a small neighborhood around the true virtual output, while the deadbeat estimate based on~\eqref{eq:linear_functional_estimator} is very noisy.
When looking at the state estimates resulting from the optimal FIE solution shown in Figure~\ref{fig:power_system_states}, we see that both the estimates of the frequency deviation $\omega$ and the mechanical power $P^M$ closely follow the ground truth states, while the phase angle/branch flow $\theta/P_{ij}$, and especially the power load $P^L$, differ largely from the ground truth values.
This can be explained by the lack of detectability of the full system state, compare our discussion in Section~\ref{sec:example_functional_detectability}.
In Figure~\ref{fig:power_system_comp_time}, we additionally show the time\footnote{The time was taken on a laptop with 12-core Intel i7 processor and 32 GB of memory.} required to solve the FIE problem depending on the time step, i.e., the number of available measurements.
The plot shows a linear relationship between computational time and number of measurements, with still a comparably small time of below $80\mathrm{ms}$ for $t=150$.
In summary, functional detectability of the system and our FIE approach allow us to obtain stable estimates of the desired virtual output (here: total power load), while, with the given measurements, it is not possible to obtain a stable overall state estimate.
\section{Conclusion}\label{sec:conclusion}
In many applications, the full state of a system is not detectable, and thus no stable state estimator can be designed.
Such problems can be approached by employing a functional estimator to compute an estimate of a (typically lower dimensional) function of the system state.
In this paper, we presented a general analysis and design framework for nonlinear functional estimation in the context of nonlinear time-varying systems subject to process and measurement noise.
We introduced a suitable characterization of nonlinear functional detectability and showed that it is a necessary and sufficient condition for the existence of an input-to-output stable functional estimator.
Thereby, the proposed functional estimator takes the form of a full information estimator.
We discussed the relation of the derived theory to existing results from functional estimation, state-norm estimation, and state estimation.
We showed practical applicability using the example of estimating the total power load of a power network, where the state is not detectable.
The design of computationally efficient nonlinear functional estimators for this general system class remains an interesting topic for future research.
First steps in this direction are recently obtained in~\cite{Muntwiler2023b}, by introducing a moving horizon estimation approach for the special case of state and parameter estimation.

\bibliographystyle{plain}        
\bibliography{bib_functional}           

\begin{thebibliography}{10}

\bibitem{Alamir2021}
Mazen Alamir.
\newblock Partial extended observability certification and optimal design of
  moving-horizon estimators.
\newblock {\em IEEE Transactions on Automatic Control}, 67(7):3663--3669, 2021.

\bibitem{allan2019lyapunov}
Douglas~A. Allan and James~B. Rawlings.
\newblock {A {L}yapunov-like Function for Full Information Estimation}.
\newblock In {\em Proc. American Control Conference (ACC)}, pages 4497--4502,
  2019.

\bibitem{Allan2021a}
Douglas~A. Allan and James~B. Rawlings.
\newblock {Robust Stability of Full Information Estimation}.
\newblock {\em SIAM Journal on Control and Optimization}, 59(5):3472--3497,
  2021.

\bibitem{Allan2021}
Douglas~A. Allan, James~B. Rawlings, and Andrew~R. Teel.
\newblock {Nonlinear Detectability and Incremental Input/Output-to-State
  Stability}.
\newblock {\em SIAM J. Contr. Opt.}, 59(4):3017--3039, 2021.

\bibitem{Andersson2019}
Joel A.~E. Andersson, Joris Gillis, Greg Horn, James~B. Rawlings, and Moritz
  Diehl.
\newblock {CasADi}: a software framework for nonlinear optimization and optimal
  control.
\newblock {\em Math. Program. Comput.}, 11(1):1--36, 2019.

\bibitem{Cai2008}
Chaohong Cai and Andrew~R. Teel.
\newblock Input–output-to-state stability for discrete-time systems.
\newblock {\em Automatica}, 44:326--336, 2008.

\bibitem{Darouach2000}
Mohamed Darouach.
\newblock {Existence and design of functional observers for linear systems}.
\newblock {\em IEEE Transactions on Automatic Control}, 45(5):940--943, 2000.

\bibitem{Darouach2020}
Mohamed Darouach and Tyrone Fernando.
\newblock {On the Existence and Design of Functional Observers}.
\newblock {\em IEEE Transactions on Automatic Control}, 65(6):2751--2759, 2020.

\bibitem{Darouach2022}
Mohamed Darouach and Tyrone Fernando.
\newblock {Functional Detectability and Asymptotic Functional Observer Design}.
\newblock {\em IEEE Transactions on Automatic Control}, 2022.

\bibitem{Emami2015}
Kianoush Emami, Tyrone Fernando, Brett Nener, Hieu Trinh, and Yang Zhang.
\newblock {A functional observer based fault detection technique for dynamical
  systems}.
\newblock {\em Journal of the Franklin Institute}, 352(5):2113--2128, 2015.

\bibitem{Fairman1980}
F.~W. Fairman and R.~D. Gupta.
\newblock {Design of multifunctional reduced order observers}.
\newblock {\em International Journal of Systems Science}, 11(9):1083--1094,
  1980.

\bibitem{Fernando2010}
Tyrone~Lucius Fernando, Hieu~Minh Trinh, and Les Jennings.
\newblock {Functional observability and the design of minimum order linear
  functional observers}.
\newblock {\em IEEE Transactions on Automatic Control}, 55(5):1268--1273, 2010.

\bibitem{Kalman1960}
R.E. Kalman.
\newblock {On the general theory of control systems}.
\newblock {\em IFAC Proceedings Volumes}, 1(1):491--502, 1960.

\bibitem{knuefer2021MHE}
Sven Kn{\"u}fer and Matthias~A. M{\"u}ller.
\newblock Nonlinear full information and moving horizon estimation: Robust
  global asymptotic stability.
\newblock {\em Automatica}, 150:110603, 2023.

\bibitem{Li2016}
Na~Li, Changhong Zhao, and Lijun Chen.
\newblock {Connecting automatic generation control and economic dispatch from
  an optimization view}.
\newblock {\em IEEE Transactions on Control of Network Systems}, 3(3):254--264,
  2016.

\bibitem{Luenberger1971}
David~G. Luenberger.
\newblock {An Introduction to Observers}.
\newblock {\em IEEE Transactions on Automatic Control}, 16(6):596--602, 1971.

\bibitem{Montanari2022}
Arthur~N. Montanari, Leandro Freitas, Daniele Proverbio, and Jorge Gonçalves.
\newblock Functional observability and subspace reconstruction in nonlinear
  systems.
\newblock {\em Physical Review Research}, 4:043195, 2022.

\bibitem{Muller2012}
Matthias~A. M{\"{u}}ller and Daniel Liberzon.
\newblock {Input/output-to-state stability and state-norm estimators for
  switched nonlinear systems}.
\newblock {\em Automatica}, 48(9):2029--2039, 2012.

\bibitem{Muntwiler2023b}
Simon Muntwiler, Johannes K\"ohler, and Melanie~N. Zeilinger.
\newblock {MHE under parametric uncertainty - Robust state estimation without
  informative data}.
\newblock {\em arXiv preprint arXiv:2312.14049}, 2023.

\bibitem{Niazi2020}
Muhammad Umar~B. Niazi, Carlos Canudas-De-Wit, and Alain~Y. Kibangou.
\newblock {Average State Estimation in Large-Scale Clustered Network Systems}.
\newblock {\em IEEE Transactions on Control of Network Systems},
  7(4):1736--1745, 2020.

\bibitem{Rao2000}
Christopher~V. Rao.
\newblock {\em {Moving horizon strategies for the constrained monitoring and
  control of nonlinear discrete -time systems}}.
\newblock PhD thesis, 2000.

\bibitem{Rawlings2020}
James~B. Rawlings, David~Q. Mayne, and Moritz Diehl.
\newblock {\em {Model Predictive Control: Theory, Computation, and Design}}.
\newblock Nob Hill Publishing, 2nd edition, 2020.
\newblock 3rd printing.

\bibitem{Schiller2022}
Julian~D. Schiller, Simon Muntwiler, Johannes Köhler, Melanie~N. Zeilinger,
  and Matthias~A. Müller.
\newblock A {Lyapunov} function for robust stability of moving horizon
  estimation.
\newblock {\em IEEE Transactions on Automatic Control}, 68(12):7466--7481,
  2023.

\bibitem{Sontag1989}
Eduardo~D. Sontag.
\newblock {Smooth Stabilization Implies Coprime Factorization}.
\newblock {\em IEEE Transactions on Automatic Control}, 34(4):435--443, 1989.

\bibitem{Sontag1998}
Eduardo~D. Sontag.
\newblock {Comments on integral variants of ISS}.
\newblock {\em Systems and Control Letters}, 34(1-2):93--100, 1998.

\bibitem{Sui2011}
Dan Sui and Tor~A. Johansen.
\newblock {Moving horizon observer with regularisation for detectable systems
  without persistence of excitation}.
\newblock {\em International Journal of Control}, 84(6):1041--1054, 2011.

\bibitem{teixeira2015strategic}
Andr{\'e} Teixeira, Henrik Sandberg, and Karl~H. Johansson.
\newblock Strategic stealthy attacks: the output-to-output $\ell_2$-gain.
\newblock In {\em 2015 54th IEEE Conference on Decision and Control (CDC)},
  pages 2582--2587. IEEE, 2015.

\bibitem{Trinh2006}
H.~Trinh, Tyrone Fernando, and S.~Nahavandi.
\newblock {Partial-state observers for nonlinear systems}.
\newblock {\em IEEE Transactions on Automatic Control}, 51(11):1808--1812,
  2006.

\bibitem{Tsui1985}
Chia~Chi Tsui.
\newblock {A New Algorithm for the Design of Multifunctional Observers}.
\newblock {\em IEEE Transactions on Automatic Control}, 30(1):89--93, 1985.

\bibitem{Venkateswaran2022}
Sunjeev Venkateswaran, Benjamin~A. Wilhite, and Costas Kravaris.
\newblock {Functional observers with linear error dynamics for discrete-time
  nonlinear systems}.
\newblock {\em Automatica}, 143:110420, 2022.

\bibitem{Wang2021}
Yingchun Wang, Baopeng Zhu, Huaguang Zhang, and Wei~Xing Zheng.
\newblock {Functional observer-based finite-time adaptive ISMC for continuous
  systems with unknown nonlinear function}.
\newblock {\em Automatica}, 125:109468, 2021.

\bibitem{Waechter2005}
Andreas Wächter and Lorenz~T. Biegler.
\newblock On the implementation of an interior-point filter line-search
  algorithm for large-scale nonlinear programming.
\newblock {\em Math. Program.}, 106(1):25--57, 2005.

\bibitem{Zhao2013}
Changhong Zhao, Ufuk Topcu, Na~Li, and Steven Low.
\newblock {Power System Dynamics as Primal-Dual Algorithm for Optimal Load
  Control}.
\newblock {\em arXiv preprint arXiv:1305.0585v1}, 2013.

\end{thebibliography}

\appendix

\ifarXiv
\section{Proofs of Sections~\ref{sec:functional_estimation} and~\ref{sec:FIE}}
This appendix contains the proof of Proposition~\ref{prop:dIOOS_necessary} (Appendix~\ref{app:proof_prop_dIOOS_nec}), Proposition~\ref{prop:dIOOS_Lyap} and Corollary~\ref{cor:exponendial_dIOOS} (Appendix~\ref{app:proof_prop_dIOOS_Lyap}), Theorem~\ref{thm:FIFE_stability} (Appendix~\ref{app:proof_thm_FIE_stability}), and Corollary~\ref{cor:FIE_quad_cost} (Appendix~\ref{app:proof_cor_FIE_quad_cost}).
\subsection{Proof of Proposition~\ref{prop:dIOOS_necessary}} \label{app:proof_prop_dIOOS_nec}
This proof extends the proof idea in~\cite[Prop. 2.6]{Allan2021}, where the special case with $\phi(x_t)=x_t$ was considered.
Consider an arbitrary initial time $t_0\in\mathbb{I}_{\ge 0}$, time step $t\in\mathbb{I}_{\ge t_0}$ and two arbitrary sequences $\{\mathbf{x},\mathbf{w},\mathbf{y},\mathbf{z}\} \in \Sigma_{t_0}^{t}$ and $\{\tilde{\mathbf{x}}, \tilde{\mathbf{w}}, \tilde{\mathbf{y}},\tilde{\mathbf{z}}\} \in \Sigma_{t_0}^{t}$.
Applying the functional estimator~\eqref{eq:functional_estimator} with $\bar{x}_{t_0} = \tilde{x}_{t_0}$, $\bar{w}_j = \tilde{w}_j$, $\bar{y}_j = \tilde{y}_j$, $j \in \mathbb{I}_{[t_0,t-1]}$, results in
\begin{align}\label{eq:nec_proof_phi_2}
	\hat{z}_t = \Psi_t\left(\tilde{x}_{t_0},\{\tilde{w}_j\}_{j=t_0}^{t-1},\{\tilde{y}_j\}_{j=t_0}^{t},t_0\right).
\end{align}
Stability of the functional estimates (Definition~\ref{def:input_output_stability}) implies that $\hat{z}_t = \tilde{z}_t$, $t\in\mathbb{I}_{\geq t_0}$.
Hence, applying the stability definition~\eqref{eq:stability_max_form} to the two sequences $z,\hat{z}$, yields
\begin{align*}
	\|z_t - \tilde{z}_t\| \le& \max\{\beta_1(\|x_{t_0} - \tilde{x}_{t_0}\|,t_0), \\
	&\max_{j\in\mathbb{I}_{[t_0,t-1]}}\beta_2(\|w_j-\tilde{w}_j\|,t-j-1), \\
	&\max_{j\in\mathbb{I}_{[t_0,t-1]}}\beta_3(\|y_j-\tilde{y}_j\|,t-j-1)\},
\end{align*}
which is equivalent to~\eqref{eq:dIOOS}.
Consequently, since the initial time step $t_0$, the time step $t$, and both sequences $\{\mathbf{x},\mathbf{w},\mathbf{y},\mathbf{z}\}$ and $\{\tilde{\mathbf{x}},  \tilde{\mathbf{w}}, \tilde{\mathbf{y}}, \tilde{\mathbf{z}}\}$ were arbitrary, the system is $\delta$-IOOS according to Definition~\ref{def:dIOOS}. \vspace*{-8mm}\begin{flushright}\qed\end{flushright}

\subsection{Proof of Proposition~\ref{prop:dIOOS_Lyap} and Corollary~\ref{cor:exponendial_dIOOS}}\label{app:proof_prop_dIOOS_Lyap}

This proof extends the results in~\cite{Allan2021,allan2019lyapunov}, where the special case with $\phi(x)=x$ was considered.
The proof is split in two parts: In Part I, we show the sufficient direction following similar steps as in~\cite[Proposition 5]{allan2019lyapunov}. In Part II, we show the necessary direction following similar steps as in~ \cite[Theorem 3.2]{Allan2021}.
Corollary~\ref{cor:exponendial_dIOOS} follows directly by considering exponential $\delta$-IOOS. \\
\textbf{Part I:} For all initial times $t_0\in\mathbb{I}_{\ge 0}$ and time steps $t\in\mathbb{I}_{\ge t_0}$, consider two arbitrary sequences both satisfying the system dynamics~\eqref{eq:sys} and the constraints~\eqref{eq:constraints}, i.e., $\{\mathbf{x,w,y,z}\}\in\Sigma_{t_0}^t$ and $\{\tilde{\mathbf{x}}, \tilde{\mathbf{w}}, \tilde{\mathbf{y}},\tilde{\mathbf{z}}\}\in\Sigma_{t_0}^t$. Starting with the initial conditions $x_{t_0}$ and $\tilde{x}_{t_0}$ at time $t_0$, and applying~\eqref{eq:IOOS_Lyap_2} $t-t_0$ times, we obtain
\begin{align}
	W_\delta(&x_t,\tilde{x}_t,t) \le \eta^{t-t_0}W_\delta(x_{t_0},\tilde{x}_{t_0},t_0) \nonumber \\
	&+ \sum_{j=t_0}^{t-1}\eta^{t-j-1}\left(\sigma_{\mathrm{w}}(\|w_j-\tilde{w}_j\|)+\sigma_{\mathrm{y}}(\|y_j-\tilde{y}_j\|)\right) \label{eq:Lyap_decrease_t_times} \\
	\stackrel{\eqref{eq:IOOS_Lyap_1}}{\le}&\eta^{t-t_0}\alpha_2(\|x_{t_0}-\tilde{x}_{t_0}\|) \nonumber\\
	& + \sum_{j=t_0}^{t-1}\eta^{t-j-1}\left(\sigma_{\mathrm{w}}(\|w_j-\tilde{w}_j\|)+\sigma_{\mathrm{y}}(\|y_j-\tilde{y}_j\|)\right). \nonumber
\end{align}
This sum-based bound also implies the following bound involving maximization (cf., e.g.,~\cite[Cor. 1]{Schiller2022}):
\begin{align*}
	W_\delta(x_t,&\tilde{x}_t,t) \le \max\left\{\vphantom{\frac{4}{1-\sqrt{\eta}}}3 \eta^{t-t_0} \alpha_2(\|x_{t_0}-\tilde{x}_{t_0}\|), \right. \\
	&\left. \max_{j\in\mathbb{I}_{[0,t-t_0-1]}}\frac{3}{1-\sqrt{\eta}}\sqrt{\eta}^{j}\sigma_{\mathrm{w}}(\|w_{t-j-1}-\tilde{w}_{t-j-1}\|), \right. \\
	&\left. \max_{j\in\mathbb{I}_{[0,t-t_0-1]}}\frac{3}{1-\sqrt{\eta}}\sqrt{\eta}^{j}\sigma_{\mathrm{y}}(\|y_{t-j-1}-\tilde{y}_{t-j-1}\|)  \right\}.
\end{align*}
Finally, applying the lower bound in~\eqref{eq:IOOS_Lyap_1} and substituting $i=t-j-1$ results in
\begin{align*}
	\|z_t-\tilde{z}_t&\| \le \max\left\{\vphantom{\frac{4}{1-\sqrt{\eta}}}\alpha_1^{-1}\left(3 \eta^{t-t_0} \alpha_2(\|x_{t_0}-\tilde{x}_{t_0}\|)\right), \right. \\
	&\left. \max_{i\in\mathbb{I}_{[t_0,t-1]}}\alpha_1^{-1}\left(\frac{3}{1-\sqrt{\eta}}\sqrt{\eta}^{t-i-1}\sigma_{\mathrm{w}}(\|w_{i}-\tilde{w}_{i}\|) \right), \right. \\
	&\left. \max_{i\in\mathbb{I}_{[t_0,t-1]}}\alpha_1^{-1}\left(\frac{3}{1-\sqrt{\eta}}\sqrt{\eta}^{t-i-1}\sigma_{\mathrm{y}}(\|y_{i}-\tilde{y}_{i}\|) \right) \right\},
\end{align*}
which corresponds to Equation~\eqref{eq:dIOOS}. \\
\textbf{Part II:} 	
Using~\cite[Prop. 7]{Sontag1998}, there exist functions $\alpha,\alpha_{\mathrm{x}},\alpha_{\mathrm{w}},\alpha_{\mathrm{y}}\in\mathcal{K}_\infty$, such that for all $k\in\mathbb{I}_{\ge 0}$, $s\ge 0$, we have
\begin{align}
	\alpha(\beta_{\mathrm{x}}(s,k))\le&\lambda^{k}\alpha_{\mathrm{x}}(s), \label{eq:beta_x_bound} \\
	\alpha(\beta_{\mathrm{w}}(s,k))\le&\lambda^{k}\alpha_{\mathrm{w}}(s), \label{eq:beta_w_bound} \\
	\alpha(\beta_{\mathrm{y}}(s,k))\le&\lambda^{k}\alpha_{\mathrm{y}}(s). \label{eq:beta_y_bound}
\end{align}
with $\lambda\coloneqq e^{-1}<1$ and $\beta_{\mathrm{x}},\beta_{\mathrm{w}},\beta_{\mathrm{y}}\in\mathcal{KL}$ from Definition~\ref{def:dIOOS}.
Applying $\alpha(\cdot)$ to both sides of the $\delta$-IOOS Condition~\eqref{eq:dIOOS} and using the fact that $\alpha(\max\{a,b\}) = \max\{\alpha(a),\alpha(b)\}$, we have
\begin{align}
	\alpha(\|z_t - \tilde{z}_t \|) \le& \max\left\{\vphantom{\max_{j\in\mathbb{I}_{[0,t-1]}}}\alpha\left(\beta_{\mathrm{x}}(\|x_{t_0} - \tilde{x}_{t_0}\|,t-t_0)\right) \right., \nonumber \\ 
	&\max_{j\in\mathbb{I}_{[t_0,t-1]}}\alpha\left(\beta_{\mathrm{w}}(\|w_j-\tilde{w}_j\|,t-j-1)\right), \nonumber \\
	&\left. \max_{j\in\mathbb{I}_{[t_0,t-1]}}\alpha\left(\beta_{\mathrm{y}}(\|y_j-\tilde{y}_j\|,t-j-1)\right)\right\}, \nonumber
\end{align}
for all $t_0\in\mathbb{I}_{\ge 0}$, $t\in\mathbb{I}_{\ge t_0}$, and any sequences $\{\mathbf{x},\mathbf{w},\mathbf{y},\mathbf{z}\}\in \Sigma_{t_0}^{t}$ and $\{\tilde{\mathbf{x}}, \tilde{\mathbf{w}}, \tilde{\mathbf{y}}, \tilde{\mathbf{z}}\}\in \Sigma_{t_0}^{t}$.
Additionally, using $\max\{a,b\}\le a + b$ for $a,b\ge 0$ and applying the bounds~\eqref{eq:beta_x_bound}-\eqref{eq:beta_y_bound}, we have
\begin{align}
	\alpha(\|&´z_t - \tilde{z}_t \|) \le \lambda^{t-t_0}\alpha_{\mathrm{x}}(\|x_{t_0} - \tilde{x}_{t_0}\|) \label{eq:dIOOS_Lyap_bound}\\
	&+ \sum_{j=t_0}^{t-1} \lambda^{t-j-1} \left( \alpha_{\mathrm{w}}(\|w_j-\tilde{w}_j\|) + \alpha_{\mathrm{y}}(\|y_j-\tilde{y}_j\|)\right). \nonumber
\end{align}
We now define a candidate $\delta$-IOOS Lyapunov function for two arbitrary states $x,\tilde{x}\in\mathbb{R}^{n_{\mathrm{x}}}$.
For $(x,\tilde{x})\notin\mathbb{X}\times\mathbb{X}$ and any $t\in\mathbb{I}_{\ge 0}$, we define $W_{\delta}(x,\tilde{x},t)=0$. For $(x,\tilde{x})\in\mathbb{X}\times\mathbb{X}$ and any $t\in\mathbb{I}_{\ge 0}$, we define
\begin{align}
	W_{\delta}(&x,\tilde{x},t) \nonumber \\
	\coloneqq& \sup_{t'\in\mathbb{I}_{\ge t},\{\mathbf{x},\mathbf{w},\mathbf{y},\mathbf{z}\}\in \Sigma_{t}^{t'},\{\tilde{\mathbf{x}}, \tilde{\mathbf{w}}, \tilde{\mathbf{y}},\tilde{\mathbf{z}}\}\in \Sigma_{t}^{t'}, x_{t} = x, \tilde{x}_{t}=\tilde{x}} \nonumber \\
	& \sqrt{\lambda}^{-(t'-t)}\left[\vphantom{\sum_{j=0}^{t-1}}\alpha\left(\|z_{t'} - \tilde{z}_{t'}\|\right) \right. \label{eq:Lyap_candidate} \\
	& \left. - \sum_{j=t}^{t'-1} \lambda^{t'-j-1} \left( \alpha_{\mathrm{w}}(\|w_j-\tilde{w}_j\|) + \alpha_{\mathrm{y}}(\|y_j-\tilde{y}_j\|)\right)\right]. \nonumber
\end{align}
In the following, we show that the candidate $W_\delta(x,\tilde{x},t)$ in~\eqref{eq:Lyap_candidate} satisfies~\eqref{eq:IOOS_Lyap_1} and~\eqref{eq:IOOS_Lyap_2} for any $(x,\tilde{x})\in\mathbb{X}\times\mathbb{X}$ and $t\in\mathbb{I}_{\ge 0}$.
Using the bound~\eqref{eq:dIOOS_Lyap_bound} in the candidate $\delta$-IOOS Lyapunov function~\eqref{eq:Lyap_candidate} and for simplicity omitting the arguments below the supremum operation, we obtain
\begin{align*}
	W_{\delta}(x,\tilde{x},t) \le& \sup \sqrt{\lambda}^{(t' -t)}\alpha_{\mathrm{x}}\left(\|x_{t} - \tilde{x}_{t}\| \right), \\
	=& \alpha_{\mathrm{x}}\left(\|x - \tilde{x}\| \right),
\end{align*}
where we used the property that $\sqrt{\lambda}^{(t' -t)}\le 1$ for all $t'\in\mathbb{I}_{\ge t}$.
This shows the upper bound in~\eqref{eq:IOOS_Lyap_1} with $\alpha_2 \coloneqq \alpha_{\mathrm{x}}$.
Furthermore, we lower bound the supremum in~\eqref{eq:Lyap_candidate} using the feasible solution $t'=t$, resulting in
\begin{align*}
	W_{\delta}(x,\tilde{x},t) \ge \lambda^0 \alpha\left( \|z_{t} - \tilde{z}_{t} \|\right) = \alpha\left(\|z - \tilde{z} \|\right),
\end{align*}
which shows the lower bound in~\eqref{eq:IOOS_Lyap_1} with $\alpha_1\coloneqq\alpha$.
Finally, to show the decrease condition~\eqref{eq:IOOS_Lyap_2}, consider an arbitrary time step $t\in\mathbb{I}_{\ge 0}$ and arbitrary $\{x,w,y,z\}\in\mathbb{X}\times\mathbb{W}\times\mathbb{Y}\times\mathbb{Z},\{\tilde{x},\tilde{w},\tilde{y},\tilde{z}\}\in\mathbb{X}\times\mathbb{W}\times\mathbb{Y}\times\mathbb{Z}$, where $y=h(x,w,t)$, $\tilde{y}=h(\tilde{x},\tilde{w},t)$, $z=\phi(x)$, and $\tilde{z}=\phi(\tilde{x})$, with $x^+=f(x,w,t)$ and $\tilde{x}^+=f(\tilde{x},\tilde{w},t)$.
In case $(x^+,\tilde{x}^+)\notin\mathbb{X}\times\mathbb{X}$, it follows by definition that $W_\delta(x^+,\tilde{x}^+)=0$ and consequently~\eqref{eq:IOOS_Lyap_2} is satisfied.
In the following, we show that the candidate Lyapunov function~\eqref{eq:Lyap_candidate} also satisfies~\eqref{eq:IOOS_Lyap_2} for $(x^+,\tilde{x}^+)\in\mathbb{X}\times\mathbb{X}$.
We note that for every $\epsilon > 0$ there exist $t^* \in \mathbb{I}_{\ge t+1}$, $\{\mathbf{x}^*,\mathbf{w}^*,\mathbf{y}^*,\mathbf{z}^*\}\in\Sigma_{t+1}^{t^*},\ \{\tilde{\mathbf{x}}^*, \tilde{\mathbf{w}}^*, \tilde{\mathbf{y}}^*, \tilde{\mathbf{z}}^*\}\in\Sigma_{t+1}^{t^*}$ with $x_{t+1}^*=x^+$ and $\tilde{x}_{t+1}^*=\tilde{x}^+$, such that
\begin{align*}
	&W_{\delta}(x^+,\tilde{x}^+,t+1) \\
	&\stackrel{\eqref{eq:Lyap_candidate}}{\le} \epsilon + \sqrt{\lambda}^{-(t^*-t-1)}\left[\vphantom{\sum_{j=1}^{t^*}}\alpha\left(\|\phi(x_{t^*}^*) - \phi(\tilde{x}_{t^*}^*)\|\right)  \right. \\
	&\left. - \sum_{j=t+1}^{t^*-1} \lambda^{t^*-j-1}\left(\alpha_{\mathrm{w}}\left(\|w_{j}^* - \tilde{w}_{j}^*\|\right) + \alpha_{\mathrm{y}}\left(\|y_{j}^*-\tilde{y}_{j}^*\|\right) \right) \right]\\
	&= \epsilon +  \sqrt{\lambda} \left(\sqrt{\lambda}^{-(t^*-t)}\left[\vphantom{\sum_{j=1}^{t^*}}\alpha\left(\|\phi(x_{t^*}') - \phi(\tilde{x}_{t^*}')\|\right)  \right. \right. \\
	&- \sum_{j=t}^{t^*-1} \lambda^{t^*-j-1}\left(\alpha_{\mathrm{w}}\left(\|w_{j}' - \tilde{w}_{j}'\|\right) + \alpha_{\mathrm{y}}\left(\|y_{j}'-\tilde{y}_{j}'\|\right) \right) \\
	&\left. \left. + \lambda^{t^*-t-1}\alpha_{\mathrm{w}}\left(\|w_t' - \tilde{w}_t'\|\right) + \lambda^{t^*-t-1}\alpha_{\mathrm{y}}\left(\|y_t' - \tilde{y}_t'\|\right) \vphantom{\sum_{j=1}^{t^*}}\right] \right),
\end{align*}
where we introduced $x_{j}' = x_{j}^*,\ \tilde{x}_{j}'=\tilde{x}_{j}^*$, $w_{j}' = w_j^*,\ \tilde{w}_{j}' = \tilde{w}_j^*$, $y_{j}' = y_j^*,\ \tilde{y}_{j}' = \tilde{y}_j^*$, $z_{j}'=z_j^*$, $\tilde{z}_{j}'=\tilde{z}_j'$ for all $j \in \mathbb{I}_{[t+1,t^*]}$, $x_t'=x,\ \tilde{x}_t' = \tilde{x}$, $z_t' = z$, $\tilde{z}_t'=\tilde{z}$, and added and subtracted the terms depending on $w_t'=w,\ \tilde{w}_t' = \tilde{w}, y_t' = y,\ \tilde{y}_t' = \tilde{y}$.
Since the sequences $\{\mathbf{x}',\mathbf{w}',\mathbf{y}',\mathbf{z}'\}\in\Sigma_{t}^{t^*}$ and $\{\tilde{\mathbf{x}}', \tilde{\mathbf{w}}', \tilde{\mathbf{y}}', \tilde{\mathbf{z}}'\}\in\Sigma_t^{t^*}$ are a feasible candidate solution for the optimization problem to obtain $W_\delta(x,\tilde{x},t)$ in~\eqref{eq:Lyap_candidate}, they result in a lower bound of the supremum.
It follows that
\begin{align*}
	W_{\delta}(x^+,\tilde{x}^+&,t+1) \\
	\stackrel{\eqref{eq:Lyap_candidate}}{\le}& \epsilon + \sqrt{\lambda}W_{\delta}(x, \tilde{x},t) \\ 
	&+ \sqrt{\lambda}^{(t^*-t-1)}\left(\alpha_{\mathrm{w}}\left(\|w - \tilde{w}\|\right) + \alpha_{\mathrm{y}}\left(\|y - \tilde{y}\|\right)\right) \\
	\le & \epsilon + \sqrt{\lambda}W_{\delta}(x, \tilde{x},t) \\ 
	&+ \alpha_{\mathrm{w}}\left(\|w - \tilde{w}\|\right) + \alpha_{\mathrm{y}}\left(\|y - \tilde{y}\|\right).
\end{align*}
Because all terms depending on $t^*$, $\{\mathbf{x}^*,\mathbf{w}^*,\mathbf{y}^*,\mathbf{z}^*\}$, and $\{\tilde{\mathbf{x}}^*, \tilde{\mathbf{w}}^*, \tilde{\mathbf{y}}^*, \tilde{\mathbf{z}}^*\}$ were removed, $\epsilon$ is arbitrary and we can choose $\epsilon \rightarrow 0$ to obtain
\begin{align*}
	W_{\delta}(x^+,\tilde{x}^+,t+1) \le& \sqrt{\lambda}W_{\delta}(x,\tilde{x},t) \\
	&+ \alpha_{\mathrm{w}}\left(\|w - \tilde{w}\|\right) + \alpha_{\mathrm{y}}\left(\|y - \tilde{y}\|\right),
\end{align*}
which shows~\eqref{eq:IOOS_Lyap_2} for all $t\in\mathbb{I}_{\ge 0}$ with $\sigma_{\mathrm{w}} = \alpha_{\mathrm{w}}$, $\sigma_{\mathrm{y}}=\alpha_{\mathrm{y}}$, and $\eta=\sqrt{\lambda}$, which concludes the proof. \vspace*{-8mm}\begin{flushright}\qed\end{flushright}

\subsection{Proof of Theorem~\ref{thm:FIFE_stability}}\label{app:proof_thm_FIE_stability}
This proof extends the results in~\cite[Prop.~2, Cor.~2]{Schiller2022},\cite{knuefer2021MHE}, which consider FIE in the special case $\phi(x)=x$.
First, note that the initial time step $t_0$ and corresponding initial estimate $\bar{x}_{t_0}$ can be chosen arbitrary in the FIE, as required for Definition~\ref{def:input_output_stability}.
Given optimality of the solution of~\eqref{eq:FIE_IOOS}, we have that the optimal objective~\eqref{eq:FIE_IOOS_cost} is upper-bounded by the objective~\eqref{eq:FIE_objective} with the true, and hence feasible, sequences, i.e.,
\begin{align}
	&V_{\mathrm{FIE}}(\hat{x}_{t_0|t}^*,\hat{w}_{\cdot|t}^*,\hat{y}_{\cdot|t}^*,t) \le \eta^{t}\alpha_2\left(2\|x_{t_0}-\bar{x}_{t_0}\|\right) \label{eq:FIE_value_upperbound} \\
	&+ \sum_{j=1}^{t-t_0}\eta^{j-1}\left(\sigma_{\mathrm{w}}(2\|w_{t-j}-\bar{w}_{t-j}\|)+\sigma_{\mathrm{y}}(2\|y_{t-j}-\bar{y}_{t-j}\|)\right), \nonumber
\end{align}
for all $t\in\mathbb{I}_{\ge t_0}$.
Applying the bound~\eqref{eq:IOOS_Lyap_2} on $W_{\delta}(\hat{x}_t,x_t,t)$ for $t-t_0$ times, as in Part~I of the proof of Proposition~\ref{prop:dIOOS_Lyap}, we have
\begin{align}\label{eq:FIE_Lyap_decrease}
	&W_{\delta}(\hat{x}_t,x_t,t) \stackrel{\eqref{eq:IOOS_Lyap_1},\eqref{eq:Lyap_decrease_t_times}}{\le} \eta^{t-t_0} \alpha_2(\|\hat{x}_{t_0|t}^*-x_{t_0}\|) \\
	& + \sum_{j=1}^{t-t_0} \eta^{j-1}\left(\sigma_{\mathrm{w}}(\|\hat{w}_{t-j|t}^* - w_{t-j}\|) + \sigma_{\mathrm{y}}(\|\hat{y}_{t-j|t}^* - y_{t-j}\|)\right). \nonumber
\end{align}
Using the weak triangular inequality of $\mathcal{K}$-functions~\cite{Sontag1989}, we have that
\begin{align*}
	\alpha_2(\|\hat{x}_{t_0|t}^*-x_{t_0}\|) &\le \alpha_2(\|\hat{x}_{t_0|t}^*-\bar{x}_{t_0}\|+\|x_{t_0}-\bar{x}_{t_0}\|) \\
	&\le \alpha_2(2\|\hat{x}_{t_0|t}^*-\bar{x}_{t_0}\|) + \alpha_2(2\|x_{t_0}-\bar{x}_{t_0}\|).
\end{align*}
Similarly, we have that
\begin{align*}
	\sigma_{\mathrm{w}}(\|\hat{w}_{t-j|t}^* - w_{t-j}\|) \le& \sigma_{\mathrm{w}}(2\|\hat{w}_{t-j|t}^*-\bar{w}_{t-j}\|) \\
	&+ \sigma_{\mathrm{w}}(2\|w_{t-j}-\bar{w}_{t-j}\|), \\
	\sigma_{\mathrm{y}}(\|\hat{y}_{t-j|t}^* - y_{t-j}\|) \le& \sigma_{\mathrm{y}}(2\|\hat{y}_{t-j|t}^*-\bar{y}_{t-j}\|) \\
	&+ \sigma_{\mathrm{y}}(2\|y_{t-j}-\bar{y}_{t-j}\|).
\end{align*}
Inserting the above three inequalities in~\eqref{eq:FIE_Lyap_decrease}, we arrive at
\begin{align*}
	&W_{\delta}(\hat{x}_t,x_t,t) \le \eta^{t} \alpha_2(2\|x_{t_0} - \bar{x}_{t_0}\|) + V_{\mathrm{FIE}}(\hat{x}_{t_0|t}^*,\hat{w}_{\cdot|t}^*,\hat{y}_{\cdot|t}^*,t) \\
	& +\sum_{j=1}^{t-t_0}\eta^{j-1} \left(\sigma_{\mathrm{w}}(2\|w_{t-j}-\bar{w}_{t-j}\|) + \sigma_{\mathrm{y}}(2\|y_{t-j}-\bar{y}_{t-j}\|) \right) \\
	&\stackrel{\eqref{eq:FIE_value_upperbound}}{\le} 2\eta^{t} \alpha_2(2\|x_{t_0}-\bar{x}_{t_0}\|) \\
	&+ 2\sum_{j=1}^{t-t_0}\eta^{j-1} \left( \sigma_{\mathrm{w}}(2\|w_{t-j}-\bar{w}_{t-j}\|) + \sigma_{\mathrm{y}}(2\|y_{t-j}-\bar{y}_{t-j}\|) \right).
\end{align*}
Applying the lower bound in~\eqref{eq:IOOS_Lyap_1} and transforming the above sum-based bound into a maximum-based formulation, as similarly done in the proof of Proposition~\ref{prop:dIOOS_Lyap}, results in 
\begin{align*}
	\|\hat{z}_t&-z_t\| \le \max\left\{\vphantom{\frac{6}{1-\sqrt{\eta}}}\alpha_1^{-1}\left(6 \eta^t \alpha_2(2\|x_{t_0}-\bar{x}_{t_0} \|)\right), \right. \\
	& \max_{j\in\mathbb{I}_{[t_0,t-1]}}\alpha_1^{-1}\left(\frac{6}{1-\sqrt{\eta}}\sqrt{\eta}^{t-j-1}\sigma_{\mathrm{w}}(2\|w_{j}-\bar{w}_{j}\|) \right), \\
	&\left. \max_{j\in\mathbb{I}_{[t_0,t-1]}}\alpha_1^{-1}\left(\frac{6}{1-\sqrt{\eta}}\sqrt{\eta}^{t-j-1}\sigma_{\mathrm{y}}(2\|y_{j}-\bar{y}_{j} \|) \right) \right\}.
\end{align*}
The above bound shows that the FIE is incrementally input-to-output stable according to Definition~\ref{def:input_output_stability}. \vspace*{-8mm}\begin{flushright}\qed\end{flushright}

\subsection{Proof of Corollary~\ref{cor:FIE_quad_cost}}\label{app:proof_cor_FIE_quad_cost}
As in the proof of Theorem~\ref{thm:FIFE_stability}, note that the initial time step $t_0$ and corresponding initial estimate $\bar{x}_{t_0}$ can be chosen arbitrary in the FIE.
Due to the enforced constraints in~\eqref{eq:FIE_IOOS}, the resulting optimal sequence of estimates satisfies the system dynamics~\eqref{eq:sys} and constraints~\eqref{eq:constraints}.
We can therefore apply the definition of exponential functional detectability~\eqref{eq:dIOOS} to both, the true and estimated sequences.
Squaring both sides results in
\begin{align*}
	\|z_t - \hat{z}_t \|^2 \le &\max\left\{C_{\mathrm{x}}^2 \lambda_{\mathrm{x}}^{2(t-t_0)}\|x_{t_0} - \hat{x}_{t_0|t}^*\|^2, \vphantom{\max_{j\in\mathbb{I}_{[0,t-1]}}} \right. \nonumber \\ 
	&\max_{j\in\mathbb{I}_{[t_0,t-1]}}C_{\mathrm{w}}^2 \lambda_{\mathrm{w}}^{2(t-j-1)}\|w_j-\hat{w}_{j|t}^*\|^2, \nonumber \\
	&\left. \max_{j\in\mathbb{I}_{[t_0,t-1]}}C_{\mathrm{y}}^2 \lambda_{\mathrm{y}}^{2(t-j-1)}\|y_j-\hat{y}_{j|t}^*\|^2\right\} \nonumber \\
	\le & C_{\mathrm{x}}^2 \eta^{2(t-t_0)}\|x_{t_0} - \hat{x}_{t_0|t}^*\|^2 \\
	&+ \sum_{j=t_0}^{t-1}C_{\mathrm{w}}^2 \eta^{2(t-j-1)}\|w_j-\hat{w}_{j|t}^*\|^2 \\
	&+ \sum_{j=t_0}^{t-1}C_{\mathrm{y}}^2 \eta^{2(t-j-1)}\|y_j-\hat{y}_{j|t}^*\|^2,
\end{align*}
where we used $\eta^2 \ge \max\{\lambda_{\mathrm{x}}^2, \lambda_{\mathrm{w}}^2, \lambda_{\mathrm{y}}^2\}$.
For any matrix $P\succ 0$ we have that
\begin{align}
	\lambda_{\min}(P)\|x\|^2 \le \|x\|_P^2 \le \lambda_{\max}(P)\|x\|^2. \label{eq:pd_norm_bound}
\end{align}
For any $P,\ Q,\ R\succ 0$ it follows that
\begin{align}
	\|z_t - \hat{z}_t \|^2 \le & C^2 \eta^{2(t-t_0)}\|x_{t_0} - \hat{x}_{t_0|t}^*\|_{P}^2 \nonumber \\
	&+ \sum_{j=t_0}^{t-1}C^2 \eta^{2(t-j-1)}\|w_j-\hat{w}_{j|t}^*\|_{Q}^2 \label{eq:cor_proof_1} \\
	&+ \sum_{j=t_0}^{t-1}C^2 \eta^{2(t-j-1)}\|y_j-\hat{y}_{j|t}^*\|_{R}^2, \nonumber
\end{align}
where we introduced 
\begin{align*}
	C^2 \coloneqq \max\left\{\nicefrac{C_{\mathrm{x}}^2}{\lambda_{\min}(P)},\ \nicefrac{C_{\mathrm{w}}^2}{\lambda_{\min}(Q)},\ \nicefrac{C_{\mathrm{y}}^2}{\lambda_{\min}(R)}\right\}.
\end{align*}
By Cauchy-Schwarz and Young's inequality we have that
\begin{align*}
	\|x_{t_0} - \hat{x}_{t_0|t}^* \|_{P}^2 \le&\ 2 \|x_{t_0} - \bar{x}_{t_0} \|_{P}^2 + 2\|\bar{x}_{t_0} - \hat{x}_{t_0|t}^* \|_{P_2}^2, \\
	\|w_j - \hat{w}_{j|t}^* \|_Q^2 \le&\ 2 \|w_j-\bar{w}_j\|_Q^2 + 2 \|\bar{w}_j - \hat{w}_{j|t}^*\|_Q^2, \\
	\|y_j - \hat{y}_{j|t}^* \|_Q^2 \le&\ 2 \|y_j-\bar{y}_j\|_Q^2 + 2 \|\bar{y}_j-\hat{y}_{j|t}^*\|_R^2.
\end{align*}
Inserting those three inequalities in~\eqref{eq:cor_proof_1}, as similarly done in the proof of Theorem~\ref{thm:FIFE_stability}, we have that
\begin{align*}
	\|z_t& - \hat{z}_t \|^2 \le \\
	&C^2 \left[ 2\eta^{2(t-t_0)}\left(\|x_{t_0} - \bar{x}_{t_0} \|_{P}^2 + \|\bar{x}_{t_0} - \hat{x}_{t_0|t}^* \|_{P}^2\right)  \vphantom{\sum_{j=0}^{t-1}}\right. \\
	&+ 2\sum_{j=t_0}^{t-1}\eta^{2(t-j-1)}\left(\|w_j-\bar{w}_j\|_Q^2+\|\bar{w}_j - \hat{w}_{j|t}^*\|_Q^2\right) \\
	&+ \left. 2\sum_{j=t_0}^{t-1} \eta^{2(t-j-1)}\left(\|y_j-\bar{y}_j\|_{R}^2+\|\bar{y}_j-\hat{y}_{j|t}^*\|_{R}^2\right) \right] \\
	\stackrel{\eqref{eq:FIE_value_upperbound}}{\le} & C^2 \left[4 \eta^{2(t-t_0)} \|x_{t_0} - \bar{x}_{t_0} \|_{P}^2 \vphantom{\sum_{j=}^{t-1}}\right. \\
	& \left. + 4 \sum_{j=t_0}^{t-1}\eta^{2(t-j-1)}\left(\|\bar{w}_j-w_j\|_Q^2 + \|\bar{y}_j - y_j \|_R^2\right)\right] \\
	\stackrel{\eqref{eq:pd_norm_bound}}{\le} & 4 C^2 \lambda_{\max}(P)\eta^{2(t-t_0)} \|x_{t_0} - \bar{x}_{t_0} \|^2 \\
	& + 4 C^2 \lambda_{\max}(Q) \sum_{j=t_0}^{t-1}\eta^{2(t-j-1)}\|w_j-\bar{w}_j\|^2 \\
	& + 4 C^2 \lambda_{\max}(R) \sum_{j=t_0}^{t-1}\eta^{2(t-j-1)}\|y_j-\bar{y}_j\|^2.
\end{align*}
Taking the square root on both sides and using that $\sqrt{a+b}\leq \sqrt{a}+\sqrt{b}$ for $a,b\ge 0$ results in
\begin{align*}
	\|z_t - \hat{z}_t¨ \| \le& 2 C \sqrt{\lambda_{\max}(P)}\eta^{t-t_0} \|x_{t_0} - \bar{x}_{t_0} \| \\
	& + 2 C \sqrt{\lambda_{\max}(Q)} \sum_{j=t_0}^{t-1}\eta^{(t-j-1)}\|w_j-\bar{w}_j\| \\
	& + 2 C \sqrt{\lambda_{\max}(R)} \sum_{j=t_0}^{t-1}\eta^{(t-j-1)}\|y_j-\bar{y}_j\|.
\end{align*}
Finally, transforming the above sum-based bound into a formulation involving maximization, as similarly done in the proof of Proposition~\ref{prop:dIOOS_Lyap}, results in
\begin{align*}
	\|\hat{z}_t&-z_t\| \le 6C\max\left\{\vphantom{\frac{6}{1-\sqrt{\eta}}} \sqrt{\lambda_{\max}(P)}\sqrt{\eta}^{t-t_0} \|x_{t_0} - \bar{x}_{t_0} \|, \right. \\
	& \max_{j\in\mathbb{I}_{[t_0,t-1]}}\frac{\sqrt{\lambda_{\max}(Q)}}{1-\sqrt{\eta}}\sqrt{\eta}^{t-j-1}2\|w_{j}-\bar{w}_{j}\|, \\
	&\left. \max_{j\in\mathbb{I}_{[t_0,t-1]}}\frac{\sqrt{\lambda_{\max}(R)}}{1-\sqrt{\eta}}\sqrt{\eta}^{t-j-1}\|y_{j}-\bar{y}_{j}\| \right\},
\end{align*}%
which shows that Definition~\ref{def:input_output_stability} is satisfied. \vspace*{-8mm}\begin{flushright}\qed\end{flushright}
\section{Derivation of $\delta$-IOOS Lyapunov Function for Linear Systems}\label{app:lin_functional}
In this section, we show that $W_\delta (x,\tilde{x})$ in~\eqref{eq:dIOSS_lin} is a $\delta$-IOOS Lyapunov function according to Definition~\ref{def:IOOS_Lyap} for the linear system~\eqref{eq:sys_lin}, assuming Condition~\eqref{eq:lin_nec_cond} holds.
For this purpose, we show in the following that~\eqref{eq:dIOSS_lin} satisfies~\eqref{eq:IOOS_Lyap}.

The upper bound in~\eqref{eq:IOOS_Lyap_1} is trivially satisfied since $W_\delta(x,\tilde{x})$ in~\eqref{eq:dIOSS_lin} is quadratic in $x-\tilde{x}$.
Additionally, we have that
\begin{align*}
	\|z-\tilde{z}\|^2 \stackrel{\eqref{eq:sys_lin}}{=}&\ \|L(x-\tilde{x})\|^2 \stackrel{\eqref{eq:lin_nec_cond_2}}{=}\ \|P_\xi T(x-\tilde{x})\|^2 \\
	\le& \frac{\lambda_{\max}(P_\xi^\top P_\xi)}{\lambda_{\min}(P)}\underbrace{\|T(x-\tilde{x})\|_P^2}_{=W_\delta(x,\tilde{x})},
\end{align*}
which implies the lower bound~\eqref{eq:IOOS_Lyap_1}.
Note that this lower bound shows that, in fact, the $\delta$-IOOS Lyapunov function is naturally only a positive semi-definite function, and not a positive definite function as in the case of state estimation.

To show that~\eqref{eq:dIOSS_lin} satisfies the decrease condition in~\eqref{eq:IOOS_Lyap_2}, consider tuples $\{x,w,y,z\}\ \in\ \mathbb{X}\times\mathbb{W}\times\mathbb{Y}\times\mathbb{Z}$, and $\{\tilde{x},\tilde{w},\tilde{y},\tilde{z}\}\ \in\ \mathbb{X}\times\mathbb{W}\times\mathbb{Y}\times\mathbb{Z}$, where $y=Cx+Dw$, $\tilde{y}=C\tilde{x}+D\tilde{w}$, $z=Lx$, and $\tilde{z}=L\tilde{x}$.
Using~\eqref{eq:sys_lin} and~\eqref{eq:lin_nec_cond_1}, we have that
\begin{align}\label{eq:Tx_plus}
	T(x^+-\tilde{x}^+) =& NT(x-\tilde{x}) \\
	&+ (TB-JD)(w-\tilde{w}) + J(y-\tilde{y}). \nonumber
\end{align}
Using Cauchy-Schwarz and Young's inequality with a sufficiently small $\epsilon > 0$ yields
\begin{align*}
	W_\delta(x^+,\tilde{x}^+) \stackrel{\eqref{eq:dIOSS_lin}}{=}&\ \|T(x^+ - \tilde{x}^+) \|_P^2 \\
	\stackrel{\eqref{eq:Tx_plus},\eqref{eq:Lyap_eq}}{\le}&\ \underbrace{(1+\epsilon) \rho}_{\coloneqq \eta} W_\delta(x,\tilde{x}) \\
	&+ C_1\|y-\tilde{y}\|^2 + C_2\|w-\tilde{w}\|^2,
\end{align*}
with $\eta < 1$ and 
\begin{align*}
	C_1 =&\ \frac{2(1+\epsilon)}{\epsilon}\lambda_{\max}(J^\top P J), \\
	C_2 =&\ \frac{2(1+\epsilon)}{\epsilon}\lambda_{\max}((TB-JD)^\top P(TB-JD)),
\end{align*}
which shows that~\eqref{eq:dIOSS_lin} is a $\delta$-IOOS Lyapunov function for system~\eqref{eq:sys_lin}.
In case the current measurement is used in the functional estimator~\eqref{eq:linear_functional_estimator} (as noted in Footnote~\ref{ft:current_measurement}), the $\delta$-IOOS Lyapunov function changes to
\begin{align*}
	W_\delta(x,\tilde{x}) = \| T(x-\tilde{x}) \|_P^2 + \| C(x-\tilde{x})\|^2.
\end{align*}
\section{Functional Detectability of Power System}\label{app:example_derivation}
In this section, we derive Equation~\eqref{eq:linear_functional_estimator}.
We start by summing the discretized and noisy form of the frequency dynamics~\eqref{eq:power_sys_freq} for all buses $i\in\mathcal{N}$, obtaining
\begin{align*}
	\sum_{i\in\mathcal{N}}\frac{M_i}{\Delta t}&\left([\omega_{t+1}]_i-[\omega_t]_i - [w_t]_{N+i}\right) \\
	&= -\sum_{i\in\mathcal{N}}\left(D_i[\omega_t]_i-[P_t^M]_i+[P_t^L]_i+[\Delta P_t]_i \right).
\end{align*}
From the definition of $[\Delta P_t]_i$ in~\eqref{eq:delta_P_i}, we have that
\begin{align*}
	\sum_{i\in\mathcal{N}}[\Delta P_t]_i = 0,
\end{align*}
because each term $P_{ij}$ appears twice, once with positive and once with negative sign.
From the measurement function~\eqref{eq:example_meas} we have that $[\omega_t]_i = [y_t]_i - [w_t]_{4N+i}$ and $[P_t^M]_i = [y_t]_{i+N} - [w_t]_{5N+i}$ for all $i\in\mathcal{N}$.
Reordering the terms and using~\eqref{eq:example_total_load} we obtain
\begin{align*}
	z_t =& \sum_{i\in\mathcal{N}} \left( \frac{M_i}{\Delta t}[w_t]_{N+i} +  [y_t]_{N+i} - [w_t]_{5N+i} \right. \\
	&- D_i([y_t]_i-[w_t]_{4N+i}) \\
	&\left. - \frac{M_i}{\Delta t}([y_{t+1}]_i - [w_{t+1}]_{4N+i} - [y_t]_i + [w_t]_{4N+i}) \right),
\end{align*}
where the right-hand side only depends linearly on process and measurement noise, and output measurements.
Using the discretized version of the power load dynamics in~\eqref{eq:power_sys} subject to process noise we have
\begin{align*}
	z_t = z_{t-2} + \sum_{i\in\mathcal{N}}\left([w_{t-1}]_{i+2N}+[w_{t-2}]_{i+2N}\right).
\end{align*}
Combining the two equations above results in~\eqref{eq:linear_functional_estimator}.
\fi

\end{document}